\UseRawInputEncoding
\documentclass[superscriptaddress, twocolumn, amsmath, amssymb, aps,pra, notitlepage,longbibliography]{revtex4-2} %preview twocolumn
\usepackage{graphicx,graphics,epsfig,subfigure,times,bm,bbm,amssymb,amsmath,amsfonts,amsthm,mathrsfs,MnSymbol,physics}
\usepackage[matrix,frame,arrow]{xypic}
\usepackage[normalem]{ulem}
\usepackage{slashed}
\usepackage{dcolumn}
\usepackage{tabularx}
\usepackage{makecell}
\usepackage{amsopn}
\usepackage{color}
\usepackage[usenames,dvipsnames,svgnames,table]{xcolor}
\usepackage[english]{babel}
\usepackage{verbatim}
\definecolor{darkblue}{rgb}{0.0,0.0,0.3}
\usepackage[colorlinks=true,
            linkcolor=red,
            urlcolor= darkblue,
            citecolor=blue]{hyperref}

\newcommand{\bea}{\begin{eqnarray}}
\newcommand{\eea}{\end{eqnarray}}

\begin{document}
\title{Anomalous flow in correlated quantum systems: No-go result and multiple-charge scenario}

\author{Rui Guan}
\address{QianWeiChang College, Shanghai University, Shanghai, 200444, China}
\author{Junjie Liu}
\email{jj\_liu@shu.edu.cn}
\affiliation{Institute for Quantum Science and Technology, Shanghai Key Laboratory of High Temperature Superconductors, Department of Physics, International Center of Quantum and Molecular Structures, Shanghai University, Shanghai, 200444, China}
\address{QianWeiChang College, Shanghai University, Shanghai, 200444, China}

\begin{abstract}
Correlated quantum systems can exhibit thermodynamic behaviors that defy classical expectations, with anomalous energy flow (AEF) against temperature gradients serving as a paradigmatic example. While AEF has been shown to arise from the consumption of initial quantum correlations, little is known about whether AEF can occur without correlation depletion, or if analogous anomalous transport exists for conserved quantities--dubbed charges--other than energy. Here, we develop a general global-local thermodynamic approach to describe charge exchange between arbitrary correlated quantum systems. For energy-conserving systems, we analytically rule out AEF in initially uncorrelated states, even with the involvement of quantum catalysts, thereby complementing existing studies. In contrast, in systems with multiple conserved charges, we uncover a mechanism for AEF that requires no initial correlations but is instead induced by a drag effect from normal flows of non-energy charges. Furthermore, by treating all conserved charges on equal footing, we generalize AEF to a broader concept of anomalous charge flow, applicable to any conserved charge. We confirm theoretical expectations with numerical examples. These findings deepen our understanding of nonequilibrium quantum thermodynamics and open new avenues for controlling transport phenomena in correlated quantum systems.
\end{abstract}

\date{\today}
\maketitle

\section{Introduction}
Quantum thermodynamics ~\cite{Anders.16.CP,Goold.16.JPA,binder2018thermodynamics} stands at the intersection of two fundamental pillars of modern physics: thermodynamics and quantum mechanics, offering both a conceptual extension of celebrated classical thermodynamics and a practical foundation for understanding the energetic aspects of future quantum technologies~\cite{Auffeves.22.PRXQ}. At microscopic scales, quantum thermodynamics introduces novel phenomena by exploiting quantum resources such as quantum coherence~\cite{Streltsov.17.RMP}, quantum correlations~\cite{Gallego.14.NJP,Bera.17.NC,Llobet.15.PRX,JiW.22.PRL} and quantum entanglement~\cite{Hilt.09.PRA,Brunner.14.PRE,Jennings.10.PRE,Alhambra.19.PRL} to name just a few. These features enable quantum thermodynamic processes that are impossible in the classical regime and fundamentally alter the limits and behaviors of traditional thermodynamics.

A particularly striking manifestation that highlights the distinctions between quantum and classical thermodynamics is anomalous energy flow (AEF)~\cite{Partovi.08.PRE,Jennings.10.PRE,Jevtic.12.PRL}, where energy can spontaneously flow from a colder subsystem to a hotter one--a phenomenon that appears to contradict the {\it classical} Clausius statement, hence it is termed anomalous. Over the past decades, significant theoretical progress has been made in understanding AEF~\cite{Dijkstra.10.PRL,Semin.12.PRA,Chaudhry.13.PRA,Zhang.15.SR,Jevtic.15.PRE,Ahmadi.21.SR,Colla.22.NJP,Henao.18.PRE,Latune.19.PRR,Medina.20.PRA,Xian.20.PRR,Zicari.20.PRR,Pusuluk.21.PRR,Gnezdilov.23.PRA,Bartosik.24.PRL,Ma.25.PRR,Comar.24.A,Mallik.24.A}, with a recent experimental confirmation~\cite{Micadei.19.NC} demonstrating its practical relevance. 
%This counterintuitive effect not only compels a reexamination of the microscopic foundations of the second law, but also provides a perfect arena for understanding the delicate interplay among quantum coherence, correlations, and thermodynamic irreversibility. 

Despite remarkable progress, the current understanding of AEF has predominantly been achieved by considering initial correlations between systems, where the reversal of energy flow is mediated by consuming these initial correlations~\cite{Dijkstra.10.PRL,Semin.12.PRA,Chaudhry.13.PRA,Zhang.15.SR,Jevtic.15.PRE,Ahmadi.21.SR,Colla.22.NJP,Henao.18.PRE,Latune.19.PRR,Medina.20.PRA,Xian.20.PRR,Zicari.20.PRR,Pusuluk.21.PRR,Bartosik.24.PRL,Ma.25.PRR,Micadei.19.NC}. While this mechanism successfully demonstrates the unique role of quantum correlations in thermodynamics, it fundamentally limits the scope of AEF due to the stringent requirements for preparing correlated initial states. In practice, this constraint has restricted experimental realizations to simple quantum systems (e.g., two-qubit setups~\cite{Micadei.19.NC}), as preparing correlated initial states in more complex composite quantum systems presents formidable experimental challenges~\cite{Bartosik.24.PRL,Ma.25.PRR,Pocklington.25.PRL,Lloyd.25.PRXQ}. This limitation highlights the need for alternative mechanisms that could enable AEF. However, it remains unclear if, and under what circumstances, AEF can manifest without consuming correlations during the energy exchange process. More broadly, quantum many-body systems with multiple conserved quantities have received increasing attention in quantum thermodynamics, sparking ongoing discussions on topics including quantum thermalization~\cite{Halpern.16.NC,Guryanova.16.NC,Lostaglio.17.NJP,Halpern.20.PRE,Kranzl.23.PRXQ,Murthy.23.PRL,Majidy.24.NC}, the refinement of thermodynamic concepts~\cite{Manzano.22.PRXQ,Upadhyaya.24.PRXQ} and entanglement enhancement~\cite{Majidy.23.PRB} among others; For more developments, we refer readers to a recent comprehensive review~\cite{Majidy.23.NRP}. Hence, one also wonders whether the notion of anomalous flow extends to other conserved quantities besides energy. For later convenience, we dub conserved quantities charges. 

In this work, we introduce a global-local thermodynamic approach to describe generic closed quantum systems with intra-system correlations, particularly including those lacking well-defined thermodynamic affinities, such as temperature, during their nonequilibrium evolution; The terminology ``global-local" will be clarified following Eq. (\ref{eq:local}), where its precise meaning becomes apparent. Inspired by an information-theoretic formulation of the second law of thermodynamics \cite{Esposito.10.NJP,Ma.21.A} in system-bath configurations, this global-local approach provides a versatile tool for analyzing exchange processes of arbitrary charges between subsystems of correlated quantum systems with or without quantum catalysts. 

Applying this approach to a conventional bipartite setting with energy conservation~\cite{Partovi.08.PRE,Jennings.10.PRE,Jevtic.12.PRL,Medina.20.PRA,Bartosik.24.PRL,Ma.25.PRR}, we establish fundamental limitations on AEF. Crucially, we consider initially uncorrelated states where correlations develop dynamically--a scenario that explicitly precludes the consumption of initial correlations, thereby going beyond the conventional mechanism of AEF. We first analytically prove the impossibility of AEF in such bipartite quantum systems. While previous studies have suggested this result~\cite{Jennings.10.PRE,Jevtic.12.PRL,Micadei.19.NC,Bartosik.24.PRL}, we note that Refs.~\cite{Jennings.10.PRE,Jevtic.12.PRL} relied on a free energy definition based solely on initial temperatures--a notion that may break down at finite times when subsystem temperatures cease to be well-defined thermodynamically. Furthermore, Refs.~\cite{Micadei.19.NC,Bartosik.24.PRL} argued this result for initially uncorrelated systems by analyzing expressions explicitly derived for initially correlated systems. Our proof provides an alternative yet more rigorous foundation for this no-go result. We then extend this analysis to settings with quantum catalyst~\cite{Bartosik.24.RMP,Junior.25.PRL}. Although quantum catalyst can enhance AEF in initially correlated systems~\cite{Bartosik.24.PRL}, we demonstrate that it cannot reverse the direction of energy flow in initially uncorrelated systems. Thus we establish a robust no-go result for AEF that holds regardless of catalytic assistance.

We further generalize our analysis to systems with multiple conserved charges, revealing a previously unidentified mechanism for AEF. In this extended scenario, we demonstrate that the net normal flow of non-energy charges can induce AEF through a drag effect--a phenomenon fundamentally distinct from the conventional correlation-mediated mechanisms, as it operates without correlation consumption or external work input. Moreover, since our approach treats all charges equivalently, we can naturally extend the concept of AEF to a broader notion of anomalous charge flow (ACF), applicable to arbitrary charges. Within this generalized framework, AEF becomes a special instance of ACF, with both phenomena being mediated by a drag mechanism. This unified perspective certainly advances our understanding of non-equilibrium transport phenomena in correlated quantum systems, establishing a foundation for studying quantum effects in charge transport. 

We remark that ACF induced by a drag mechanism represents a distinct phenomenon in coupled transport involving multiple charges~\cite{Berdanier.19.PRL,Wang.20.PRL,Ghosh.24.A,Manzano.22.PRXQ}. Unlike a drag effect between charge currents of the same type~\cite{Berdanier.19.PRL}, the drag mechanism identified here couples distinct charge species. We also note that ACF is inherently transient, persisting only during finite time intervals due to unitary system dynamics, in contrast to steady-state coupled transport~\cite{Manzano.22.PRXQ}. In addition, the manifestation of ACF in two-charge models lies in the mutually parallel charge currents (identical sign) despite opposing thermodynamic gradients (opposite signs), as illustrated in Fig. \ref{fig:AET} (b) and Fig. \ref{fig:ACT} (b). This contrasts with the so-called inverse current in coupled transport~\cite{Wang.20.PRL,Ghosh.24.A}, which requires instead mutually parallel thermodynamic gradients across different charge species.

The structure of the paper is as follows. In Sec.~\ref{sec:no-go}, we present a global-local thermodynamic approach that allows us to rigorously prove that AEF cannot occur in initially uncorrelated systems, even with catalysts involved. In Sec.~\ref{sec:multiple}, we extend our analysis to transport setups with multiple conserved charges. We uncover a distinct mechanism for AEF due to the drag effect of a net normal flow of remaining non-energy charges. We also generalize the notion of anomalous flow to other non-energy charges and propose the concept of ACF which subsumes AEF as a special case. Additionally, we substantiate our theoretical expectations with numerical simulations. In Sec.~\ref{sec:extension}, we further extend the analysis to account for scenarios without strict charge conservation or with arbitrary initial states. Finally, we summarize our study in Sec.~\ref{sec:conclusion} with some final remarks. For the sake of clarity, we relegate lengthy derivation details to Appendix \ref{a:1} which also includes a notation table.

\section{No-go results for anomalous energy flow} \label{sec:no-go}
In this section, we analytically show that AEF cannot occur in energy-conserving bipartite systems initialized in uncorrelated states, even when assisted by quantum catalysts. We establish this no-go result by developing a global-local thermodynamic description capable of describing charge exchange processes in generic correlated quantum systems. Crucially, our approach requires only the minimal assumption of an initial product state (Nevertheless, we will infer from Sec. \ref{sec:extension} B that this minimal assumption can also be lifted in our formulism), without imposing additional constraints on the details of the system and evolution or requiring ad hoc thermodynamic definitions.

\subsection{Conventional bipartite setup: A global-local description}
We first follow conventional lead~\cite{Dijkstra.10.PRL,Semin.12.PRA,Chaudhry.13.PRA,Zhang.15.SR,Jevtic.15.PRE,Ahmadi.21.SR,Colla.22.NJP,Henao.18.PRE,Latune.19.PRR,Medina.20.PRA,Xian.20.PRR,Zicari.20.PRR,Pusuluk.21.PRR,Gnezdilov.23.PRA,Bartosik.24.PRL,Ma.25.PRR,Comar.24.A,Mallik.24.A,Micadei.19.NC} used for analyzing AEF in which one considers a closed bipartite quantum system consisting of two non-interacting (possibly finite-dimensional) subsystems $A$ and $B$ with Hamiltonian $H_A$ and $H_B$, respectively. Departing from these prior studies, we explicitly assume that the two subsystems are initially uncorrelated and in their individual thermal equilibrium states, namely, the initial state of the composite system takes a product form $\rho(0) = \gamma_{A} \otimes \gamma_{B}$ with Gibbsian states $\gamma_{X}=e^{-\beta_{X} H_X}/Z_X$ ($X=A,B$, setting $k_B=1$ and $\hbar=1$ hereafter). We stress that $\beta_{A,B}$ introduced here just denote initial inverse temperatures which can be fixed through initial state preparation~\cite{Cheng.23.PRA} and need not have a thermodynamic interpretation at later times. In fact, during the dynamical evolution, the finite-dimensional subsystems can be out of equilibrium, rendering a meaningful thermodynamic temperature ill-defined. For later convenience, we set $\beta_B<\beta_A$, that is, subsystem $B$ is initially hotter than subsystem $A$~\footnote{We do not consider the scenario of $\beta_A=\beta_B$ where the energy exchange process is completely frozen under an energy-conserving unitary transformation as considered here.}

The closed bipartite system evolves as $\rho(0)\to\rho(t)=\mathcal{U}(t)\rho(0)\mathcal{U}^{\dagger}(t)$, where the unitary transformation $\mathcal{U}(t)$ is required to conserve the total energy at all times $t\in[0,\tau]$ with $\tau$ the duration of process. We note that $\mathcal{U}$ can be realized by involving an interaction Hamiltonian that commutes with $H_0$ and is assumed to be switched on and off at the initial and final times of the process~\cite{Jennings.10.PRE,Bartosik.24.PRL}, respectively. In general, $\rho(t)$ at finite times deviates from a product form due to the generation of correlation between subsystems. We denote $\rho_{A,B}(t)$ to be the reduced states of $\rho(t)$ for subsystem $A$ and $B$, respectively. We remark that no assumptions are made on the forms of $\rho_{A,B}(t)$ at finite times.

To deal with energy exchange processes between two finite-dimensional and correlated quantum systems where thermodynamic temperature during nonequilibrium evolution cannot be properly defined~\cite{Liu.23.PRAa}, we develop a global-local thermodynamic description for such composite systems by combining thermodynamic properties of both the global composite system and its local subsystems, as will be detailed in the following. This approach adapts and extends quantum information-theoretic methods originally established for deriving the second law of thermodynamics in system-bath configurations~\cite{Esposito.10.NJP}. We remark that the global-local approach only requires well-defined initial temperatures as input parameters and is well-suited to treat finite-dimensional systems prepared in uncorrelated initial states without invoking additional assumptions, thereby providing an alternative route to bypass the traditional treatments of AEF. 

Viewing the subsystem $B$ as a finite-dimensional environment for the subsystem $A$, we adapt and extend the approach \cite{Esposito.10.NJP} to find that the change in the von Neumann entropy of subsystem A reads (see details in Appendix \ref{a:1})
\begin{equation}\label{eq:dsa}
\Delta S_A(t)~=~D[\rho(t)||\rho_A(t) \otimes \gamma_B] - \beta_B\Delta E_B(t).
\end{equation}
Here, we have defined $\Delta A(t)=A(t)-A(0)$ to denote a change in an arbitrary observable $A(t)$ throughout the study. $D[\rho_1||\rho_2]=\mathrm{Tr}[\rho_1(\ln \rho_1-\ln\rho_2)]$ is the quantum relative entropy between states $\rho_{1,2}$. $E_X(t)=\mathrm{Tr}[\rho_X(t)H_X]$ ($X=A,B$). We note that the above expression only requires an initial product state as well as an initial inverse temperature $\beta_B$ to hold regardless of the details of $H_{A,B}$ and the dynamical process, thereby applying to arbitrary forms of bipartite systems and $\mathcal{U}$. Switching the roles of subsystems $A$ and $B$, we can get a similar expression for subsystem $B$, 
\begin{equation}\label{eq:dsb}
    \Delta S_B(t)~=~D[\rho(t)||\rho_B(t) \otimes \gamma_A]-\beta_A\Delta E_A(t).
\end{equation}
We emphasize that Eqs. (\ref{eq:dsa}) and (\ref{eq:dsb}) together treat the two subsystems on an equal footing. 
%Since the expressions [cf. Eqs. (\ref{eq:dsa}) and (\ref{eq:dsb})] involve the global state, we refer to the perspective that yields them as a global one.

To appropriately define AEF, we adopt the conventional figure of merit $(\beta_A-\beta_B)\Delta E_A(t)$~\cite{Dijkstra.10.PRL,Semin.12.PRA,Chaudhry.13.PRA,Zhang.15.SR,Jevtic.15.PRE,Ahmadi.21.SR,Colla.22.NJP,Henao.18.PRE,Latune.19.PRR,Medina.20.PRA,Xian.20.PRR,Zicari.20.PRR,Pusuluk.21.PRR,Gnezdilov.23.PRA,Bartosik.24.PRL,Ma.25.PRR,Comar.24.A,Mallik.24.A,Micadei.19.NC}. This quantity serves as an unambiguous diagnostic: when $(\beta_A-\beta_B)\Delta E_A(t)<0$, the system exhibits AEF where energy transfers against the initial temperature gradient $\beta_A-\beta_B$ during the finite time interval $[0,t]$, whereas $(\beta_A-\beta_B)\Delta E_A(t)\ge0$ marks a normal energy flow along the initial thermal gradient. Combining Eqs. (\ref{eq:dsa}) and (\ref{eq:dsb}) together, we find
\begin{equation} \label{eq:global}
(\beta_A-\beta_B)\Delta E_A(t)~=~-\Delta I(t)+\sum_{i=A,B}D[\rho(t)||\rho_i(t) \otimes \gamma_{\tilde{i}}].
\end{equation}
Here, we have utilized the relations $\Delta I(t)=\sum_{i=A,B} \Delta S_i(t)$ with $I(t)=S_A(t)+S_B(t)-S(t)$ the quantum mutual information and $\Delta E_B(t)=-\Delta E_A(t)$ due to the invariant of total entropy under unitary evolution and energy conservation, respectively; $S(t)$ denotes the von Neumann entropy of the total system. We have also denoted the subscript $\tilde{i}$ to be the complement of $i$, namely, $i=A$ then $\tilde{i}=B$ and vice versa. 

From Eq. (\ref{eq:global}), we can get an equivalent expression where the quantum relative entropy just involves local states. We note that $D[\rho(t)||\rho_A(t) \otimes \gamma_B]=I(t)+D[\rho_B(t)||\gamma_B]$ \cite{Strasberg.17.PRX} (see Appendix \ref{a:1} for more details); A similar expression holds for $D[\rho(t)||\rho_B(t) \otimes \gamma_A]$. Inserting these expressions for quantum relative entropy into Eq. (\ref{eq:global}), we receive an equivalent expression for the figure of merit, 
\begin{equation}\label{eq:local}
  (\beta_A-\beta_B)\Delta E_A(t)~=~\Delta I(t)+\sum_{i=A,B}D[\rho_i(t)||\gamma_i].
\end{equation}
Here, we have utilized the fact that $I(t)=\Delta I(t)$ in initially uncorrelated systems. We note that the right-hand side of Eq. (\ref{eq:local}) has been adopted as a definition of the total entropy production~\cite{Manzano.22.PRXQ}. We stress that both Eqs. (\ref{eq:global}) and (\ref{eq:local}) represent exact formulations that are mutually convertible through appropriate transformations.  For clarity in subsequent discussions, we refer to Eqs. (\ref{eq:global}) and (\ref{eq:local}) as the global and local descriptions, respectively--a designation reflecting the characteristic states (global $\rho(t)$ versus local $\rho_i(t)$) employed in quantum relative entropy of each expression. Together, Eqs. (\ref{eq:global}) and (\ref{eq:local}) constitute the global-local description of the composite correlated system.

To analyze whether AEF can occur between initially uncorrelated systems, we reformulate Eqs. (\ref{eq:global}) and (\ref{eq:local}) as thermodynamic inequalities to align with established analysis in the field. Noting $\Delta I(t)\ge 0$ in initially uncorrelated bipartite systems and utilizing the non-negativity of quantum relative entropy, the simultaneous validity of Eqs. (\ref{eq:global}) and (\ref{eq:local}) guarantees the following single inequality to hold,
\begin{equation} \label{eq:eq_AEF_local}
(\beta_A-\beta_B)\Delta E_A(t) \geqslant \Delta I(t).
\end{equation}
We remark that the above inequality coincides with that obtained in Refs. \cite{Bartosik.24.PRL,Jennings.10.PRE,Micadei.19.NC}. However, here we obtain it from a distinct global-local approach without utilizing the notion of local free energy or directly working with initially correlated bipartite setups. The above inequality clearly implies that $(\beta_A-\beta_B)\Delta E_A(t)\ge 0$, thereby conclusively ruling out the occurrence of AEF in initially uncorrelated bipartite systems.

We emphasize that the synthesis of global and local descriptions is crucial in our theoretical analysis. First, the local description Eq. (\ref{eq:local})--which directly gives rise to the key inequality Eq. (\ref{eq:eq_AEF_local})--is derived from its global counterpart Eq. (\ref{eq:global}). Second, and most crucially, exclusive reliance on the global description in Eq. (\ref{eq:global}) yields a qualitatively different inequality,
\begin{equation}\label{eq:eq_AEF_global}
(\beta_A-\beta_B)\Delta E_A(t)~\geqslant~-\Delta I(t). 
\end{equation}
This result presents a conceptual pitfall: while the condition $(\beta_A-\beta_B)\Delta E_A(t)<0$ appears mathematically permissible (since $-\Delta I(t)\leq 0$ in initially uncorrelated systems), this interpretation directly contradicts the physically consistent conclusion obtained from Eq. (\ref{eq:eq_AEF_local}). This discrepancy underscores that the global description alone cannot reliably predict AEF, as it fails to simultaneously satisfy both global and local thermodynamic constraints. Only through their combined analysis can we obtain unambiguous conclusions about energy flow phenomena. 

This apparent paradox between ostensibly equivalent global and local descriptions when formulated as inequalities reveals several insights that are worth further emphasizing: (i) One should be cautious when dealing with loose thermodynamic bounds since they can be misleading. (ii) The transformation between global and local descriptions fundamentally depends on quantum mutual information, as evidenced by the relation $D[\rho(t)||\rho_A(t) \otimes \gamma_B] = I(t) + D[\rho_B(t)||\gamma_B]$. Crucially, the conversion from the global description [Eq. (\ref{eq:global})] to the local description [Eq. (\ref{eq:local}]) involves an information gain quantified by $2 I(t)\ge 0$. This additional information gain in the local description serves to reconcile the apparent contradiction, turning a loose global bound into a tighter local constraint that properly determines the possibility of AEF.

%======================================================
\subsection{Extending to setup with quantum catalyst}
Notably, quantum catalyst can offer new dynamical pathways to lift restrictions imposed by the unitarity of quantum mechanics \cite{Bartosik.24.RMP}. Exploiting this capability of quantum catalyst, we now explore whether AEF can emerge in the presence of quantum catalyst under conditions previously demonstrated to prohibit its occurrence. Extending the previously studied bipartite setup, we introduce an additional catalytic component $C$ characterized by the Hamiltonian $H_C$, forming an enlarged composite system with an initial product state $\rho_{\rm tot}(0) = \rho(0)\otimes\omega_C$ where $\rho(0)=\gamma_A \otimes \gamma_B$. The initial state $\omega_C$ of the catalyst should be determined self-consistently [see Eq. (\ref{eq:cata_state}) below]. The enlarged composite system undergoes a global, energy-conserving unitary evolution $\mathcal{U}$, $\rho_{\rm tot}(t)=\mathcal{U}(t)\rho_{\rm tot}(0)\mathcal{U}^{\dagger}(t)$, ensuring strict energy conservation for the total system during the evolution process with a time interval $[0,\tau]$.

For a valid catalyst, it should return to its initial state at the end of the process. This requires the catalyst state to satisfy the following self-consistent condition \cite{Bartosik.24.RMP,Bartosik.24.PRL}
\begin{equation}\label{eq:cata_state}
  \omega_C=\text{Tr}_S[\mathcal{U}(\tau)\rho_{\rm tot}(0)\mathcal{U}^{\dagger}(\tau)].
\end{equation}
Notably, we impose no restrictions on the catalyst type. Eq. (\ref{eq:cata_state}) implies a zero net energy change $\Delta E_C(\tau)=0$ with $E_C(t)=\mathrm{Tr}[H_C\rho_C(t)]$ the internal energy and $\rho_C(t)$ the reduced state of the catalyst at time $t\in[0,\tau]$. Consequently, energy flow is confined exclusively within the conventional bipartite system after a complete catalytic process, facilitating a clear determination of the direction of energy flow and thus identifying the occurrence of AEF~\cite{Bartosik.24.PRL}.

Considering a global description of the enlarged composite system $SC$, Eq. (\ref{eq:dsa}) extends to (see details in Appendix \ref{a:1})
\begin{equation} \label{eq:entropy_catalyst}
\Delta S_A(t)~=~D[\rho_{\rm tot}(t)||\rho_A(t)\otimes\gamma_B\otimes \mathrm{I}_C]-\beta_B\Delta E_B(t)+S_C(0).
\end{equation}
To arrive at the above equality, we have utilized the facts that the global unitary evolution preserves the total entropy and no initial correlations exist between the subsystems. $\mathrm{I}_C$ denotes an identity matrix of the dimension of $H_C$. $S_C(0)=-\text{Tr}[\omega_C \ln\omega_C]$ is the initial von Neumann entropy of the catalyst. Analogously, Eq. (\ref{eq:dsb}) can also be generalized to read $\Delta S_B(t)=D[\rho_{\rm tot}(t)||\rho_B(t) \otimes \gamma_A\otimes \mathrm{I}_C]-\beta_A\Delta E_A(t)+S_C(0)$. Combining these two equations for $\Delta S_{A,B}(t)$ together, we can get the following equation that generalizes Eq. (\ref{eq:global}),
\bea\label{eq:catalyst_energy_flow_equality}
(\beta_A-\beta_B)\Delta E_A(t) &=& -\Delta I_{\rm tot}(t)+S_C(t)+S_C(0)\nonumber\\
&&+D[\rho_{\rm tot}(t)||\rho_A(t) \otimes \gamma_B\otimes \mathrm{I}_C]\nonumber\\
&&+D[\rho_{\rm tot}(t)||\rho_B(t) \otimes \gamma_A\otimes \mathrm{I}_C]\nonumber\\
&&+\beta_B\Delta E_C(t).
\eea
Here, $S_C(t)$ denotes the von Neumann entropy of catalyst at finite times, we have invoked the energy conservation condition $\sum_{i=A,B,C}\Delta E_i(t)=0$ during the process to express $\Delta E_B(t)$ in terms of $\Delta E_{A,C}(t)$. We have also utilized the relation $-\sum_{i=A,B}\Delta S_i(t)=\Delta I_{\rm tot}(t)+\Delta S_C(t)$ due to the invariance of von Neumann entropy of the enlarged composite system during unitary evolution; $I_{\rm tot}(t)\equiv \sum_{i=A,B,C}S_i(t)-S_{\rm tot}(t)$ denotes the multipartite quantum mutual information~\cite{Groisman.05.PRA,Modi.12.RMP} for the enlarged composite system at time $t$ with $S_{\rm tot}(t)$ the von Neumann entropy of the enlarged composite system. Similar to the conventional bipartite setup, the above global description can be reduced to a local one by noting the relation $D[\rho_{\rm tot}(t)||\rho_i(t) \otimes \gamma_j\otimes \mathrm{I}_C]=I_{\rm tot}(t)-S_C(t) + D[\rho_j(t)||\gamma_j],\ (i,j = A,B; i\neq j)$ (see details in Appendix \ref{a:1}),
\bea\label{eq:llocal_c}
(\beta_A-\beta_B)\Delta E_A(t) &=& \Delta I_{\rm tot}(t)-\Delta S_C(t)+\beta_B\Delta E_C(t)\nonumber\\
&&+\sum_{i=A,B}D[\rho_{i}(t)||\gamma_i].
\eea
Here, we have utilized the fact that $I_{\rm tot}(0)=0$. Eq. (\ref{eq:llocal_c}) directly generalizes Eq. (\ref{eq:local}).

To analyze whether AEF can occur in the setup with a quantum catalyst, one can still resort to thermodynamic bounds on $(\beta_A-\beta_B)\Delta E_A(t)$. By exploiting the non-negativity of quantum relative entropy, Eqs. (\ref{eq:catalyst_energy_flow_equality}) and (\ref{eq:llocal_c}) together yield the following combined inequality,
\begin{equation}\label{eq:cata_inequality}
(\beta_A-\beta_B)\Delta E_A(t) ~\geqslant~ \mathrm{max}\Big\{\mathcal{L}_g(t),\mathcal{L}_l(t)\Big\}
\end{equation}
Here, $\mathcal{L}_g(t)\equiv \beta_B \Delta E_C(t)+ S_C(t)+S_C(0)-\Delta I_{\rm tot}(t)$ and $\mathcal{L}_l(t)\equiv\beta_B \Delta E_C(t)-\Delta S_C(t)+\Delta I_{\rm tot}(t)$ denote lower bounds obtained from
global description Eq. (\ref{eq:catalyst_energy_flow_equality}) and local description Eq. (\ref{eq:llocal_c}), respectively. The symbol ``$\mathrm{max}\{\mathcal{O}_1,\mathcal{O}_2\}$" picks the larger one among the two quantities $\mathcal{O}_{1,2}$. We note that the two lower bounds $\mathcal{L}_{g,l}(t)$ in Eq. (\ref{eq:cata_inequality}) differ in an amount of $2I_{\rm tot}(t)-2S_C(t)$ which originates from the global-local transformation (see Appendix \ref{a:1}). Since we just have $I_{\rm tot}(t)\le 2~\mathrm{min}\{S(t),S_C(t)\}$ \cite{Araki.70.CMP}, the contrast $2I_{\rm tot}(t)-2S_C(t)$ lacks a definite sign in general. Hence both the relative magnitude between the two lower bounds $\mathcal{L}_{g,l}(t)$ and their signs are undetermined, unlike the conventional catalyst-free setup where the relative magnitude between two lower bounds is evident such that one can just focus on a specific single inequality Eq. (\ref{eq:eq_AEF_local}) obtained from local description to analyze the possibility of AEF.

During the catalytic process with $t\in[0,\tau)$, we note that both $\Delta E_C(t)\neq 0$ and $\Delta S_C(t)\neq 0$ since the catalyst state is not constrained to be fixed, yet their signs are not specified, further rendering indefinite signs of lower bounds in Eq. (\ref{eq:cata_inequality}). This implies that AEF, as conventionally marked as $(\beta_A-\beta_B)\Delta E_A(t)<0$, can probably occur during the catalytic process. However, one should bear in mind that catalyst can exchange energy with subsystems $A$ and $B$ during the process such that $(\beta_A-\beta_B)\Delta E_A(t)<0$ may not necessarily signifies the occurrence of AEF between just subsystems $A$ and $B$. Hence, we here just focus on the whole catalytic process where the catalyst faithfully completes its catalytic role such that energy can only flow between subsystems $A$ and $B$. Considering the whole catalytic process, Eq. (\ref{eq:cata_inequality}) reduces to 
\begin{equation}\label{eq:catalyst_final}
(\beta_A-\beta_B)\Delta E_A(\tau)~\geqslant~\mathrm{max}\Big\{2 S_C(0)-\Delta I_{\rm tot}(\tau),~\Delta I_{\rm tot}(\tau)\Big\}.
\end{equation}
Here, we have utilized the facts that $\Delta E_C(\tau)=0$ and $\Delta S_C(\tau)=0$. For a correlating catalyst which can correlate with the bipartite system at the end of the process~\cite{Bartosik.24.RMP}, we have $\Delta I_{\rm tot}(\tau)> 0$ in general (noting $I_{\rm tot}(0)=0$). Under this condition, Eq. (\ref{eq:catalyst_final}) invariably ensures that $(\beta_A-\beta_B)\Delta E_A(t)>0$ regardless of the magnitude of $2 S_C(0)-\Delta I_{\rm tot}(\tau)$~\footnote{If $2 S_C(0)-\Delta I_{\rm tot}(\tau)>\Delta I_{\rm tot}(\tau)>0$, then Eq. (\ref{eq:catalyst_final}) implies that $(\beta_A-\beta_B)\Delta E_A(\tau)\geqslant 2 S_C(0)-\Delta I_{\rm tot}(\tau)>0$. Instead, if $2 S_C(0)-\Delta I_{\rm tot}(\tau)<\Delta I_{\rm tot}(\tau)$, we have $(\beta_A-\beta_B)\Delta E_A(\tau)\geqslant \Delta I_{\rm tot}(\tau)>0$ from Eq. (\ref{eq:catalyst_final}).}. Even for a strict catalyst that returns unperturbed and uncorrelated with the bipartite
system at the end of the process \cite{Bartosik.24.RMP}, we still have $\Delta I_{\rm tot}(\tau)>0$. This follows from the definition $I_{\rm tot}(\tau)=\sum_{i=A,B,C}S_i(t)-S_{\rm tot}(t)$ which can captures residual correlations between subsystems $A$ and $B$ of the bipartite system at the end of the process. Hence, we analytically prove that the energy transfer between two initially uncorrelated quantum systems remains bound by the classical thermodynamic constraints, that is, AEF cannot occur--even with the assistance of a quantum catalyst.

%===============================================================
\section{Anomalous flow in multiple-charge scenario}\label{sec:multiple}
In this section, we demonstrate that the previously established no-go result can be circumvented in transport setups involving initially uncorrelated subsystems possessing multiple conserved charges. Specifically, we identify a mechanism for AEF mediated through a drag effect, where normal flows of non-energy charges can induce AEF. Furthermore, we generalize AEF by introducing the notion of anomalous charge flow (ACF), which describes the counter-gradient transport of any charge relative to its conjugate thermodynamic affinity. This theoretical extension establishes a unified framework for understanding anomalous transport phenomena across multiple conserved charges in correlated quantum systems.

\subsection{Anomalous energy flow due to a drag effect}
\subsubsection{Theory}
Previously, we have demonstrated that AEF cannot occur in initially uncorrelated systems where only energy is conserved. Here we extend the thermodynamic framework by incorporating multiple conserved charges besides energy. This generalization allows us to investigate the potential influence of additional charges on the direction and magnitude of energy flow and explore whether it is possible to activate AEF without consuming correlations as a resource. 

We consider a generalized bipartite transport setup where each subsystem contains its own sets of charges $\{\mathcal{C}_i^j\}$ and associated affinities $\{\lambda_i^j\}$ ($j=A,B$)~\cite{Manzano.22.PRXQ,Upadhyaya.24.PRXQ}. We assume that the two subsystems are initially uncorrelated, with each prepared in its own generalized Gibbsian state~\cite{Jaynes.57.PR,Halpern.16.NC,Guryanova.16.NC,Lostaglio.17.NJP}
\begin{equation}\label{eq:ggs}
\gamma_j~=~\frac{e^{-\sum_{i=0}\lambda_i^j \mathcal{C}_i^j}}{\mathcal{Z}_j}.
\end{equation}
Here, $\mathcal{Z}_j=\text{Tr}\left[e^{-\sum_{i=0}\lambda_i^j \mathcal{C}_i^j}\right]$ denotes the partition function. We specifically identify $\mathcal{C}_0^j\equiv H_j$ to be the subsystem's Hamiltonian with $\lambda_0^j\equiv\beta_j$ corresponding to initial inverse temperatures. We have the basic commutator $[\mathcal{C}_0^{j},\mathcal{C}_{i\neq 0}^{j}]=0$ between energy and non-energy charges, while non-energy charges can be noncommuting with $[\mathcal{C}_{n}^{j},\mathcal{C}_{m}^{j}]\neq0$ ($n\neq 0, m\neq 0, n\neq m$) since Eq. (\ref{eq:ggs}) remains valid even for noncommuting charges~\cite{Halpern.16.NC,Guryanova.16.NC,Lostaglio.17.NJP}. The bipartite system evolves under a charge-conserving unitary operator $\mathcal{U}$ which satisfies
\begin{equation}\label{eq:charge-conserving unitary}
[\mathcal{U},\sum_{j=A,B}\mathcal{C}_{i}^{j}]~=~0,~\textrm{for~all}~i.
\end{equation}
The exchange of charges between two subsystems then occurs in the presence of non-equal affinities $\{\lambda_i^A\neq \lambda_i^B\}$.

By replacing the initial Gibbsian states with these generalized Gibbsian ones, Eq. (\ref{eq:dsa}) generalizes to
\begin{equation} \label{eq:entropy_multiple}
\Delta S_A(t)~=~D[\rho(t)||\rho_A(t) \otimes \gamma_B]-\sum_{i=0} \lambda_i^{B}\Delta C_i^{B}(t).
\end{equation}
Here, we have defined $C_i^{j}(t)\equiv\mathrm{Tr}[\rho_j(t)\mathcal{C}_i^j],~\forall i$ with $j=A,B$. A similar expression for $\Delta S_B(t)$ can be obtained by swapping the indices $A\leftrightarrow B$ in the above equation. Summing up these two expressions for von Neumann entropy changes, we find
\begin{equation}\label{eq:multiple_equality}
\sum_{i=0}(\lambda_i^A-\lambda_i^B)\Delta C_i^A(t) = -\Delta I(t)+\sum_{j=A,B}D[\rho(t)||\rho_j(t) \otimes \gamma_{\tilde{j}}]
\end{equation}
that generalizes Eq. (\ref{eq:global}) to account for exchange processes of multiple charges. To get the above expression, we have invoked the charge conservation condition $\sum_{j=A,B}\Delta C_i^j=0$ and utilized the quantum mutual information $I(t)$ to replace the sum $\sum_{j=A,B} \Delta S_j(t)$. A local description, $\sum_{i=0}(\lambda_i^A-\lambda_i^B)\Delta C_i^A(t) = \Delta I(t)+\sum_{j=A,B}D[\rho_j(t)||\gamma_j]$, can be obtained from Eq. (\ref{eq:multiple_equality}) by noting the relation 
$D[\rho(t)||\rho_i(t) \otimes \gamma_j]=\Delta I(t)+ D[\rho_j(t)||\gamma_j]$ with $(i,j = A,B;~i\neq j)$ (see Appendix \ref{a:1}).

We now utilize Eq. (\ref{eq:multiple_equality}) and its local counterpart to analyze whether AEF can occur in this scenario without initial correlations. To this end, we still focus on the inequalities satisfied by the figure of merit $(\beta_A-\beta_B)\Delta E_A(t)$. The global and local descriptions lead to the combined inequality, $(\beta_A-\beta_B)\Delta E_A(t)\geqslant\mathrm{max}\Big\{\mathcal{L}_g^E(t),\mathcal{L}_l^E(t)\Big\}$. Here, $\mathcal{L}_g^E(t) = -\Delta I(t) -\sum_{i \neq 0}(\lambda_i^A-\lambda_i^B) \Delta C_i^A(t)$ and $\mathcal{L}_l^E(t) = \Delta I(t) -\sum_{i \neq 0}(\lambda_i^A-\lambda_i^B) \Delta C_i^A(t)$ are lower bounds derived from the global description Eq. (\ref{eq:multiple_equality}) and its local counterpart using the non-negativity of quantum relative entropy, respectively. The superscript `$E$' in the lower bounds distinguishes them from counterparts in Eq. (\ref{eq:cata_inequality}). Noting $\Delta I(t)\ge 0$ in initially uncorrelated setups, we finally have
\begin{equation} \label{eq:multiple_combined inequality}
(\beta_A-\beta_B)\Delta E_A(t)~\geqslant~\mathcal{L}_l^E(t).
\end{equation}
Comparing Eq. (\ref{eq:multiple_combined inequality}) with Eq. (\ref{eq:eq_AEF_local}), we note that the lower bound in Eq. (\ref{eq:multiple_combined inequality}) contain an additional contribution $-\sum_{i \neq 0}(\lambda_i^A-\lambda_i^B) \Delta C_i^A(t)$ due to the inclusion of non-energy charges, rendering both the sign and magnitude of lower bound $\mathcal{L}_l^E(t)$ indefinite in general. Crucially, we can expect that AEF characterized by $(\beta_A-\beta_B)\Delta E_A(t)<0$ is possible to occur whenever a necessary condition $\mathcal{L}_l^E(t)<0$ is fulfilled.

To gain more insights, we highlight several fundamental characteristics that distinguish this AEF mechanism in multiple-charge systems from existing results. First, the AEF in multiple-charge scenarios occurs without consuming initial quantum correlations which is completely absent, a crucial departure from the conventional correlation-consuming mechanism addressed in existing studies~\cite{Dijkstra.10.PRL,Semin.12.PRA,Chaudhry.13.PRA,Zhang.15.SR,Jevtic.15.PRE,Ahmadi.21.SR,Colla.22.NJP,Henao.18.PRE,Latune.19.PRR,Medina.20.PRA,Xian.20.PRR,Zicari.20.PRR,Pusuluk.21.PRR,Gnezdilov.23.PRA,Bartosik.24.PRL,Ma.25.PRR,Comar.24.A,Mallik.24.A,Micadei.19.NC}. Second, the current AEF is not a refrigeration since it operates without external work input or consumption of other charges. Third, the necessary condition $\Delta I(t)-\sum_{i \neq 0}(\lambda_i^A-\lambda_i^B)\Delta C_i^A(t)< 0$ physically signifies that non-energy charges exhibit a net normal flow since $\sum_{i \neq 0}(\lambda_i^A-\lambda_i^B)\Delta C_i^A(t)>0$, which actively drives the AEF through a drag effect. We remark that this drag phenomenon is distinct from existing similar ones~\cite{Berdanier.19.PRL,Wang.20.PRL,Ghosh.24.A,Manzano.22.PRXQ} as we explicitly contrasted in the Introduction. The drag mechanism of AEF not only eliminates the experimentally challenging requirement of preparing initial quantum correlations, but also substantially expands the range of physical conditions under which AEF can be realized. Moreover, this condition highlights how the net thermodynamic force from non-energy charges’ normal transport can overcome information constraints to enable counter-gradient flow of energy.

%===================================================================
\subsubsection{Numerical example}
To illustrate the occurrence of AEF due to the drag effect of a net normal flow of non-energy charges, we employ a minimum bosonic system where the energy and particle number serve as charges \cite{Santos.16.PRE}. The bosonic model we consider consists of two bosonic modes with annihilation operators $a_j$ ($j=A,B$) and on-site energy $\varepsilon_j$, with its Hamiltonian reading
\begin{equation}
    H_0~=~\sum_{j=A,B} \varepsilon_j\,a_j^\dagger a_j.
\end{equation}
Here, $H_j=\varepsilon_j\,a_j^\dagger a_j$ ($j=A,B$). We take $\varepsilon_A=\varepsilon_B = \varepsilon$. From the above form of Hamiltonian, it is evident that both the local energy and particle number $\mathcal{N}_j = a_j^\dagger a_j$ of subsystems are conserved charges. For later convenience, we denote $N_j(t)=\mathrm{Tr}[\mathcal{N}_j\rho_j(t)]$ as the averaged particle number.

The charge-conserving unitary transformation $\mathcal{U}$ is realized by introducing a charge-conserving interaction between two subsystems~\cite{Bartosik.24.PRL},
\begin{equation}
   H_{\rm I}~=~-J\bigl(a_A^\dagger a_B + a_B^\dagger a_A\bigr).
\end{equation}
Here, $J$ marks the hopping strength between two bosonic modes. This interaction is turned on during the time interval $[0,t]$. It can be easily verified that $[H_0,H_{\rm I}]=0$ and $[\mathcal{N}_A+\mathcal{N}_B,H_{\rm I}]=0$. We assume that the evolution of the composite system state $\rho(t)$, governing by the Liouville–von Neumann equation $\partial\rho(t)/\partial t=-i[H_0+H_{\rm I},\rho(t)]$, during the time interval $[0,t]$ where the interaction is turned on, starts from an initial product state where each subsystem is prepared in a grand-canonical state,
\begin{equation}
  \gamma_j~=~\frac{e^{-\beta_j\,(H_j - \mu_j \mathcal{N}_j)}}{\mathcal{Z}_j}.  
\end{equation}
Here, $\beta_j$ and $\mu_j$ denote just the initial inverse temperature and chemical potential of the $j$ bosonic mode. $\mathcal{Z}_j=\Tr\bigl[e^{-\beta_j\,(H_j - \mu_j \mathcal{N}_j)}\bigr]$ denotes the partition function. Compared with Eq. (\ref{eq:ggs}), we know that $\lambda_0^j=\beta_j$ and $\lambda_1^j=-\beta_j\mu_j$.

\begin{figure}[b!]
 \centering
\includegraphics[width=1\columnwidth]{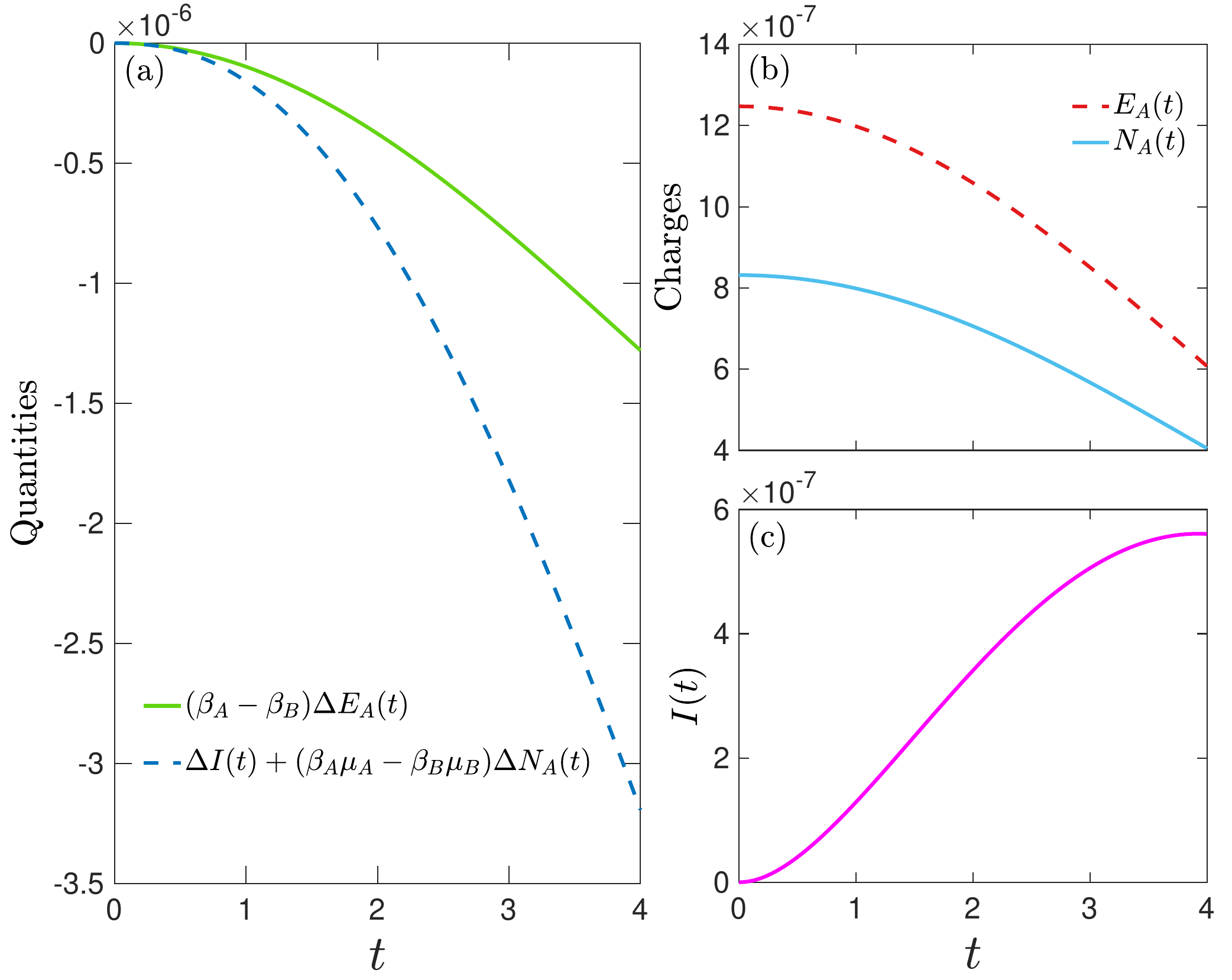} 
 \caption{Verification of drag-induced AEF: (a) Results for $(\beta_A-\beta_B)\Delta E_A(t)$ (green solid line) and $\Delta I(t)+(\beta_A\mu_A-\beta_B\mu_B)\Delta N_A(t)$ (blue dashed line) as a function of time. (b) Evolution of charges of subsystem $A$: $E_A(t)=\mathrm{Tr}[H_A\rho(t)]$ (red dashed line) and $N_A(t)=\mathrm{Tr}[\mathcal{N}_A\rho(t)]$ (blue solid line). (c) The monotonic increasing trend of quantum mutual information $I(t)$ between two bosonic systems.  Parameters are $\beta_A=20,\beta_B=18$, $\mu_A=0.8,\mu
 _B=0.4$, $\varepsilon=1.5$, $J=0.2$.}
\protect\label{fig:AET}
\end{figure}

A set of converged numerical results that demonstrates the occurrence of AEF in this bosonic system is depicted in Fig. \ref{fig:AET}. 
%==========================================
%==========================================
Fig. \ref{fig:AET} (a) provides clear evidence of AEF, demonstrated by the negative value of the figure of merit $(\beta_A-\beta_B)\Delta E_A(t)<0$. Crucially, this AEF is driven by a drag mechanism from a normal particle flow, as confirmed by the simultaneous observation of $\Delta I(t)+(\beta_A\mu_A- \beta_B\mu_B)\Delta N_A(t)<0$ in Fig. \ref{fig:AET} (a) and the charge dynamics shown in Fig. \ref{fig:AET} (b). These numerical results validate our theoretical prediction of a drag-induced AEF in multiple-charge systems. In this setup, Eq. (\ref{eq:multiple_combined inequality}) reduces to a simple inequality
\begin{equation}
(\beta_A -\beta_B)\Delta E_A(t)~\geq~\Delta I(t)+(\beta_A\mu_A - \beta_B\mu_B)\Delta N_A(t),
\end{equation}
whose validity is directly supported by Fig. \ref{fig:AET} (a). Notably, this AEF mechanism operates without correlation consumption, as evidenced by the monotonically increasing quantum mutual information $I(t)$ between the bosonic modes in Fig. \ref{fig:AET} (c). This fundamentally distinguishes our observation from conventional AEF, which requires initial correlations as an essential resource. We remark that AEF just represents a transient phenomenon since unitary evolution inevitably leads to oscillatory energy exchange between subsystems with energy eventually flowing back into subsystem $A$ at later times~\cite{Micadei.19.NC}.

%====================================================================
\subsection{Anomalous charge flow}
Building upon our demonstration of AEF induced through normal non-energy charge transport, we now establish a dual phenomenon for non-energy charges that is rooted in the symmetric treatment of all charges in Eq. (\ref{eq:multiple_equality}). This theoretical symmetry naturally suggests the notion of anomalous charge flow (ACF) which subsumes AEF as a special case. Hence, all charges can flow against their conjugate thermodynamic gradients in principle. Below, we theoretically demonstrate the existence of ACF for non-energy charges that mirrors previous AEF results, followed by numerical verification of this effect, thereby validating the concept of ACF.

%===================================================================
\subsubsection{Theory}
Rearranging Eq. (\ref{eq:multiple_equality}) and its local equivalence by isolating an arbitrary quantity $(\lambda_k^A-\lambda_k^B) \Delta C_k^A(t)$ for $k$th non-energy charge $(k\neq 0)$ as the focus of demonstrating ACF and invoking the non-negativity of the quantum relative entropy, we can obtain two lower bounds on $(\lambda_k^A-\lambda_k^B) \Delta C_k^A(t)$, namely, $\mathcal{L}_g^C(t)= -\Delta I(t) -\sum_{i \neq k}(\lambda_i^A-\lambda_i^B) \Delta C_i^A(t)$ and $\mathcal{L}_l^C(t) = \Delta I(t) -\sum_{i \neq k}(\lambda_i^A-\lambda_i^B) \Delta C_i^A(t)$ that are dual to $\mathcal{L}_g^E(t)$ and $\mathcal{L}_l^E(t)$ given above Eq. (\ref{eq:multiple_combined inequality}), respectively. By virtue of the non-negativity of quantum mutual information, it suffices to consider $\mathcal{L}_l^C(t)$ for the analysis of whether ACF can occur since $\mathcal{L}_l^C(t)\ge \mathcal{L}_g^C(t)$. We thus arrive at the following inequality which underlies our discussion of ACF
\begin{equation} \label{eq:ACT_combined inequality}
(\lambda_k^A-\lambda_k^B) \Delta C_k^A(t)~\geqslant~\mathcal{L}_l^C(t).
\end{equation}
Similar to Eq. (\ref{eq:multiple_combined inequality}), here the lower bound also lacks a definite magnitude and sign in general. Crucially, the necessary condition for the emergence of ACF is encapsulated by the requirement that $\Delta I(t) -\sum_{i \neq k}(\lambda_i^A-\lambda_i^B) \Delta C_i^A(t) <0$ which signifies a net normal flow of remaining charges besides the $k$th one, permitting $(\lambda_k^A-\lambda_k^B) \Delta C_k^A(t)$ to become negative. Physically, this means that the selected charge $C_k$ can flow against its conjugate thermodynamic gradient, in direct analogy to the drag-induced AEF phenomenon.

The concept of ACF introduced here represents a direct generalization of the established AEF framework, revealing a profound symmetry among all conserved charges. This theoretical proposal demonstrates that energy does not occupy a privileged position in nonequilibrium quantum transport phenomena--the underlying drag mechanism operates universally across all conserved charges. Crucially, similar to AEF, the realization of ACF: (i) requires no initial quantum correlations, and (ii) is fundamentally driven by cumulative normal flows of other charges, as quantitatively governed by the condition $\sum_{i \neq k}(\lambda_i^A-\lambda_i^B)\Delta C_i^A(t) > \Delta I(t) \geq 0$.

%===================================================================
\subsubsection{Numerical example}
To numerically demonstrate ACF, we consider a two-qubit model that can be related to the previous bosonic system~\footnote{In this representation, each bosonic site is associated with a spin degree of freedom, where the presence of a boson corresponds to the spin-down state, and the vacuum state corresponds to the spin-up state, see Refs. \cite{PhysRevE.94.032139,PhysRevA.109.042208} for more details.}
\begin{equation}
H_0 ~=~ -\sum_{i=A,B}\frac{\varepsilon_i}{2}\,\sigma_i^z.
\end{equation}
Here, $\varepsilon_i$ marks the energy gap of spins and $\sigma_i^z$ is the Pauli-$z$ operator at site $i$. We consider $\varepsilon_A=\varepsilon_B=\varepsilon$. Similar to the bosonic system considered above, the charge-conserving unitary transformation $\mathcal{U}$ is implemented by introducing a charge-conserving interaction between two spins
\begin{equation}
H_{\rm I}~=~ -\frac{J}{2}\bigl(\sigma_A^x\,\sigma_B^x+\sigma_A^y\,\sigma_B^y\bigr),
\end{equation}
where $J$ represents the spin–exchange strength and $\sigma_i^{x,y}$ denotes the corresponding Pauli matrices. It is straightforward to verify that the free and interaction Hamiltonian commute, $[H_0,H_{\rm I}]=0$, ensuring both the energy and the excitation number operator of each spin $\mathcal{N}_i=(\mathbb{I}-\sigma_i^z)/2$ \cite{PhysRevE.94.032139,PhysRevA.109.042208}
are conserved charges during evolution. Here, $\mathbb{I}$ denotes a $2\times 2$ identity matrix.

The time evolution of the total density matrix $\rho(t)$ is governed by the Liouville–von Neumann equation $\partial\rho(t)/\partial t=-i[H_0+H_{\rm I},\rho(t)]$ during the time interval $[0,t]$ where the interaction is turned on. We still consider an initial product state between two spins, with each subsystem prepared in a grand-canonical state~\cite{PhysRevE.94.032139,PhysRevA.109.042208},
\begin{equation}
  \gamma_i~=~\frac{e^{-\beta_i\,(H_i - \mu_i \mathcal{N}_i)}}{\mathcal{Z}_i}.  
\end{equation}
Here, $\beta_i$ and $\mu_i$ represent, respectively, the initial inverse temperature and the effective potential conjugate to the excitation number of spin $i$. In this spin setup, $\mu_j$ plays the role of an adjustable parameter controlling the initial population of excitations, analogous to the chemical potential in the previous bosonic system. Importantly, these parameters characterize only the initial equilibrium condition and should not be treated as fixed parameters during the subsequent unitary evolution. $\mathcal{Z}_i = \text{Tr}\left[\exp\left(-\beta_i (H_i - \mu_i \mathcal{N}_i)\right)\right]$ is the partition function corresponding to each spin.

\begin{figure}[thb!]
 \centering
\includegraphics[width=1\columnwidth]{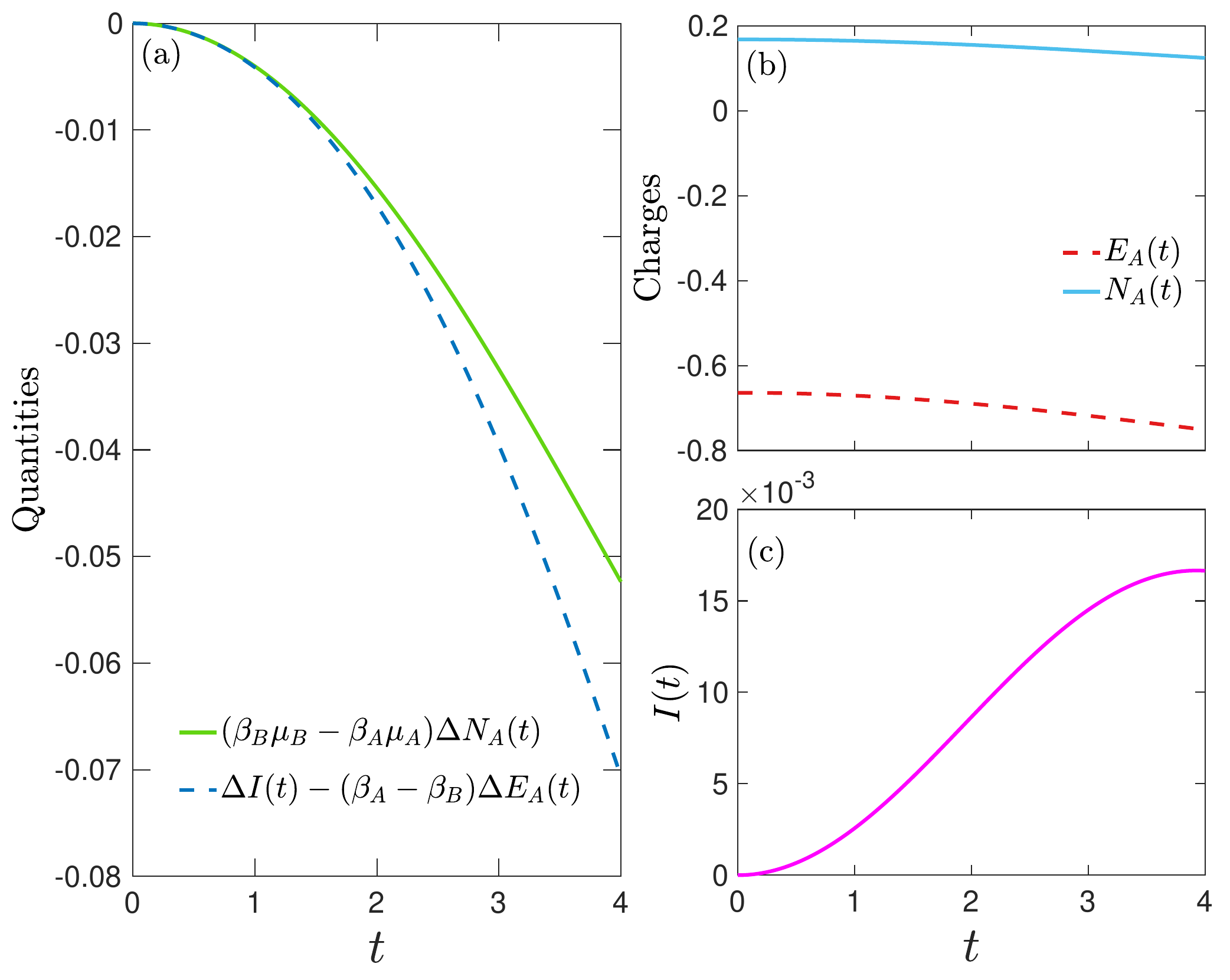} 
 \caption{Verification of ACF: (a) Results of $(\beta_B\mu_B-\beta_A\mu_A)\Delta N_A(t)$ (green solid line) and $\Delta I(t)-(\beta_A-\beta_B)\Delta E_A(t)$ (blue dashed line) as a function of time, (b) Evolution of charges of subsystem $A$: energy $E_A(t)=\mathrm{Tr}[H_A\rho(t)]$ (red dashed line) and excitation $N_A(t)=\mathrm{Tr}[\mathcal{N}_A\rho(t)]$ (blue solid line), (c) The monotonic increase of the quantum mutual information $I(t)$ between the two spins. Parameters are $\beta_A=1$, $\beta_B=2$, $\mu_A=0.4$, $\mu_B=0.8$, $\varepsilon=2$, $J=0.2$.}
\protect\label{fig:ACT}
\end{figure}

We depict a set of results in Fig. (\ref{fig:ACT}) that presents numerical evidence for the ACT in this minimum spin model.
%===============================================
%===============================================
Fig. (\ref{fig:ACT}) (a) illustrates the fundamental quantities characterizing the occurrence of ACF, where the negative value of $(\beta_B\mu_B-\beta_A\mu_A)\Delta N_A(t)$ provides direct numerical evidence for excitation flow against its thermodynamic gradient. The corresponding lower bound in Eq. (\ref{eq:ACT_combined inequality})--which simplifies to $\Delta I(t)-(\beta_A-\beta_B)\Delta E_A(t)$ in this case--similarly becomes negative, conclusively demonstrating that this ACT is drag-induced by a normal energy flow characterized by $(\beta_A-\beta_B)\Delta E_A(t)\ge 0$. The underlying mechanism is further elucidated in Fig. (\ref{fig:ACT}) (b), which tracks the time evolution of both energy $E_A(t)$ and excitation number $N_A(t)$ in subsystem $A$. The trends clearly reveal how normal energy transport drives the counter-gradient flow of excitations, providing direct visual confirmation of the drag effect. Most significantly, Fig. (\ref{fig:ACT}) (c) shows that the quantum mutual information $I(t)$ increases monotonically throughout the process. This observation confirms two crucial aspects of our mechanism: (i) the ACF occurs without correlation consumption, and (ii) the effect is fundamentally distinct from conventional correlation-driven transport phenomena. The simultaneous satisfaction of these conditions establishes drag-induced ACT as a previously unidentified class of anomalous non-equilibrium quantum transport.

%===============================================================
\section{Possible extensions}\label{sec:extension}
In this section, we examine the persistence of anomalous flow when relaxing two conventional assumptions adopted in existing studies~\cite{Dijkstra.10.PRL,Semin.12.PRA,Chaudhry.13.PRA,Zhang.15.SR,Jevtic.15.PRE,Ahmadi.21.SR,Colla.22.NJP,Henao.18.PRE,Latune.19.PRR,Medina.20.PRA,Xian.20.PRR,Zicari.20.PRR,Pusuluk.21.PRR,Gnezdilov.23.PRA,Bartosik.24.PRL,Ma.25.PRR,Comar.24.A,Mallik.24.A,Micadei.19.NC} and our previous analyses: strict charge conservation and initial local Gibbsian state or its generalized form. We limit to systems with multiple charges in the following.

\subsection{Systems without charge conservation}
In this subsection, we consider the case where charge conservation is relaxed while maintaining initial local generalized Gibbsian states for simplicity. Starting from Eq. (\ref{eq:entropy_multiple}) and its counterpart for $\Delta S_B(t)$ obtained by swapping the indices $A\leftrightarrow B$, we generalize the local reduction of Eq. (\ref{eq:multiple_equality}) to account for charge non-conservation effects that may originate from external charge sources or drains, 
\bea\label{eq:26}
    \sum_{i=0}(\lambda_i^A-\lambda_i^B)\Delta C_i^A(t) &=& \Delta I(t)+\sum_{j=A,B}D[\rho_j(t)||\gamma_j]\nonumber\\
    &&-\sum_{i=0}\lambda_i^B\sum_{j=A,B}\Delta C_i^j(t).
\eea
The final term vanishes when charge conservation $\sum_{j=A,B}\Delta C_i^j(t)=0$ is enforced. Without loss of generality, we focus on energy flow, noting that generalization of the analysis to other charges is straightforward. Building on Eq. (\ref{eq:26}), we derive the generalized form of Eq. (\ref{eq:multiple_combined inequality}),
\bea\label{eq:27}
   (\beta_A-\beta_B)\Delta E_A(t) &\ge& \Delta I(t)-\sum_{i\neq 0}(\lambda_i^A-\lambda_i^B)\Delta C_i^A(t)\nonumber\\
    &&-\sum_{i=0}\lambda_i^B\sum_{j=A,B}\Delta C_i^j(t).
\eea
The lower bound in Eq. (\ref{eq:27}) reveals that AEF remains possible when 
\begin{equation}
   \sum_{i\neq 0}(\lambda_i^A-\lambda_i^B)\Delta C_i^A(t)+\sum_{i=0}\lambda_i^B\sum_{j=A,B}\Delta C_i^j(t)~\ge~0. 
\end{equation}
This condition demonstrates that a net normal flow of non-energy charges can still drag AEF, regardless of the sign of $\sum_{j=A,B}\Delta C_i^j(t)$. Notably, when $\sum_{j=A,B}\Delta C_i^j(t)<0$, the net normal flow of non-energy charges must be sufficiently strong to compensate $\sum_{j=A,B}\Delta C_i^j(t)$ in order to observe AEF. A similar conclusion extends to ACF since we treat all charges at the equal footing.

\subsection{Systems with arbitrary initial states}
In this subsection, we turn to the scenario with nonequilibrium initial states where the local reduced states $\rho_j(0)$ ($j=A,B$) can take non-Gibbsian forms. To balance generality with simplicity, we allow for initial correlated states while enforcing charge conservation. For non-Gibbsian initial states, the initial thermodynamic affinities appeared in our previous expressions become ill-defined. To proceed, we follow Ref.~\cite{Mondal.23.A} to introduce initial local reference states of a generalized Gibbsian form for subsystems  
\begin{equation}\label{eq:reference_state}
    \gamma_j^{R}~=~\frac{e^{-\sum_{i=0}\tilde{\lambda}_i^j \mathcal{C}_i^j}}{\mathcal{Z}_j^R}.
\end{equation} 
Here, the superscript `R' indicates the reference state. $\mathcal{Z}_j^R=\text{Tr}[e^{-\sum_{i=0}\tilde{\lambda}_i^j \mathcal{C}_i^j}]$ denotes the corresponding partition function. $\{\tilde{\lambda}_i^j\}$ are reference thermodynamic affinities determined by the following set of constraints
\begin{equation}
    \text{Tr}[\rho_j(0)\mathcal{C}_i^j]~=~\text{Tr}[\gamma_j^{R} \mathcal{C}_i^j],\quad \forall i.
\end{equation}
We stress that $\{\tilde{\lambda}_i^j\}$ only acquire proper thermodynamic meaning when the subsystem is indeed thermalized at the initial time. When this is the case, we have $\gamma_j^R=\gamma_j$ in Eq. (\ref{eq:ggs}) and our generalized framework presented below naturally reduces to the previously analyzed scenarios. In the following, we will first generalize key theoretical expressions and examine the challenges in defining a thermodynamically meaningful anomalous charge flow for local nonequilibrium initial states. Subsequently, we will revisit a special case of local equilibrium initial states by using reduced forms of generalized expressions, addressing a theoretical gap left by Ref.~\cite{Bartosik.24.PRL}.

\subsubsection{Quantum catalyst-free systems}
We start from considering quantum catalyst-free bipartite systems. In this scenario, the change in von Neumann entropy of subsystem $A$ can be expressed as (see details in Appendix \ref{a:1})
\bea\label{eq:31}
    \Delta S_A(t)~&=&~D[\rho(t)||\rho_A(t) \otimes \gamma_B^{R}]-D[\rho_B(0)||\gamma_B^{R}]\nonumber\\
    &&-\sum_{i=0} \tilde{\lambda}_i^{B}\Delta C_i^{B}(t)-I(0).
\eea
Here, $I(0)$ denotes the initial quantum correlations between two subsystems. We remark that a nonvanishing quantum relative entropy $D[\rho_B(0)||\gamma_B^{R}]$ highlights the non-Gibbsian nature of the actual initial state $\rho_B(0)$. When requiring $I(0)=0$ and preparing actual initial generalized Gibbsian states $\rho_j(0)=\gamma_j$, Eq. (\ref{eq:31}) reduces to Eq. (\ref{eq:entropy_multiple}) in the previous section. Similarly, by swapping the indices $A\leftrightarrow B$, we can obtain an expression for $\Delta S_B(t)$. 

Summing up generalized expressions for $\Delta S_{A,B}(t)$, we can generalize Eq. (\ref{eq:multiple_equality}) to obtain
\bea\label{eq:extension_equality_g}
   \sum_{i=0}(\tilde{\lambda}_i^A-\tilde{\lambda}_i^B)\Delta C_i^A(t) &=& \sum_{j=A,B}\Big(D[\rho(t)||\rho_j(t) \otimes \gamma_{\tilde{j}}^{R}]\nonumber\\
   &&-D[\rho_j(0)||\gamma_j^{R}]\Big)-I(t)\nonumber\\
   &&-I(0).
\eea
Here, we have invoked the replacement $\sum_{j=A,B}\Delta S_j(t)=\Delta I(t)$ due to the invariance of total von Neumann entropy $S(t)=S(0)$. We have also utilized the charge conservation condition $\Delta C_i^A(t)=-\Delta C_i^B(t)$. We regard Eq. (\ref{eq:extension_equality_g}) as a generalized global description of charge exchange process between two initially correlated subsystems. Utilizing the relation $D[\rho(t)||\rho_A(t) \otimes \gamma_B^R] = I(t) + D[\rho_B(t)||\gamma_B^R]$ (see Appendix \ref{a:1}), we can obtain a generalized local description
\bea\label{eq:extension_equality_l}
    \sum_{i=0}(\tilde{\lambda}_i^A-\tilde{\lambda}_i^B)\Delta C_i^A(t) &=&\Delta I(t)+\sum_{j=A,B}\Big(D[\rho_j(t)||\gamma_j^{R}]\nonumber\\
    &&-D[\rho_j(0)||\gamma_j^{R}]\Big).
\eea
We remark that this formula can also be derived from a modified Landauer's principle~\cite{Mondal.23.A}. 

We note that defining ACF through Eqs. (\ref{eq:extension_equality_g}) and (\ref{eq:extension_equality_l}) may lack direct thermodynamic interpretation, as the gradients $\tilde{\lambda}_i^A-\tilde{\lambda}_i^B$ are reference-based. Nevertheless, by interpreting these reference affinities as effective parameters, one can define an effective ACF characterized by $(\tilde{\lambda}_i^A-\tilde{\lambda}_i^B)\Delta C_i^A(t) < 0$. To examine the possibility of this effective phenomenon, we can transform Eqs. (\ref{eq:extension_equality_g}) and (\ref{eq:extension_equality_l}) into inequalities by noting $\sum_{j=A,B}D[\rho(t)||\rho_j(t) \otimes \gamma_{\tilde{j}}^{R}]\ge 0$ and $\sum_{j=A,B}D[\rho_j(t)||\gamma_j^{R}]\ge 0$, respectively. Further noting that $\Delta I(t)\ge -[I(t)+I(0)]$, we only need to consider the following single inequality 
\begin{equation}\label{eq:34}
    \sum_{i=0}(\tilde{\lambda}_i^A-\tilde{\lambda}_i^B)\Delta C_i^A(t)~\ge~\Delta I(t)-\sum_{j=A,B}D[\rho_j(0)||\gamma_j^{R}].
\end{equation}
From the above lower bound, we immediately infer that an effective ACF can indeed occur since the conventional mechanism of correlation consumption with $\Delta I(t)<0$ and the drag mechanism we have identified can act cooperatively to increase the negative magnitude of the lower bound.

\subsubsection{Quantum catalyst-involved systems}
We then turn to setups with quantum catalyst. We assume that the bipartite system is initially uncorrelated with the quantum catalyst such that the total initial state takes a product form, $\rho_{\rm tot}(0)=\rho(0)\otimes \omega_C$ (Recall that we allow for correlated states $\rho(0)$). The quantum catalyst's state $\omega_C$ should be determined in a self-consistent manner [cf. Eq. (\ref{eq:cata_state})]. Under unitary evolution that conserves all charges for the composite system, the change in von Neumann entropy for subsystem $A$ reads (see details in Appendix \ref{a:1})
\bea\label{eq:35}
    \Delta S_A(t) &=& D[\rho_{\rm tot}(t)||\rho_A(t) \otimes \gamma_B^R\otimes \mathrm{I}_C]+S_C(0)-I_{\rm tot}(0) \nonumber \\
   &&-D[\rho_B(0)||\gamma_B^R]-\sum_{i=0}\tilde{\lambda}_i^B\Delta C_i^B(t).
\eea
A similar expression for $\Delta S_B(t)$ is obtained by switching the indices $A\leftrightarrow B$. Considering the sum of $\Delta S_A(t)+\Delta S_B(t)$ and invoking the charge conservation condition $\sum_{j=A,B,C}\Delta C_i^j(t)=0$, we can obtain the following equality including the effect of quantum catalyst
\bea\label{eq:36}
\sum_{i=0}(\tilde{\lambda}_i^A-\tilde{\lambda}_i^B)\Delta C_i^A(t) &=& D[\rho_{\rm tot}(t)||\rho_A(t) \otimes \gamma_B^R\otimes \mathrm{I}_C]\nonumber\\
&&+D[\rho_{\rm tot}(t)||\rho_B(t) \otimes \gamma_A^R\otimes \mathrm{I}_C]\nonumber\\
&&+S_C(t)+S_C(0)-I_{\rm tot}(t)-I_{\rm tot}(0)\nonumber\\
&&-\sum_{j=A,B}D[\rho_j(0)||\gamma_j^{R}]\nonumber\\
&&+\sum_{i=0}\tilde{\lambda}_i^B\Delta C_i^C(t).
\eea
We note that the above equation directly generalizes both Eqs. (\ref{eq:catalyst_energy_flow_equality}) and (\ref{eq:extension_equality_g}). We emphasize that the quantum catalyst need not support multiple charges like subsystems $A$ and $B$--it may conserve only energy. Consequently, the sum $\sum_{i=0}\tilde{\lambda}_i^B\Delta C_i^C(t)$ just includes nonzero contributions from charges of quantum catalyst. Invoking the global-local transformation for quantum relative entropy (see Appendix \ref{a:1}), Eq. (\ref{eq:36}) can be transformed into
\bea\label{eq:37}
\sum_{i=0}(\tilde{\lambda}_i^A-\tilde{\lambda}_i^B)\Delta C_i^A(t) &=&
\Delta I_{\rm tot}(t)-\Delta S_C(t)+\sum_{i=0}\tilde{\lambda}_i^B\Delta C_i^C(t)\nonumber\\
&&+\sum_{j=A,B}\Big(D[\rho_j(t)||\gamma_j^{R}]\nonumber\\
&&-D[\rho_j(0)||\gamma_j^{R}]\Big).
\eea
It is evident that Eq. (\ref{eq:37}) generalizes Eqs. (\ref{eq:llocal_c}) and (\ref{eq:extension_equality_l}). 

Considering a complete catalytic process of a time interval $[0,\tau]$ with $S_C(\tau)=S_C(0)$ and $\Delta C_i^C(\tau)=0$, Eqs. (\ref{eq:36}) and (\ref{eq:37}) together imply the following inequality by using the non-negativity of quantum relative entropy terms involving finite-time states,
\begin{widetext}
\begin{equation}\label{eq:38}
\sum_{i=0}(\tilde{\lambda}_i^A-\tilde{\lambda}_i^B)\Delta C_i^A(\tau)~\ge~\left\{\begin{array}{ll}
2S_C(0)-I_{\rm tot}(\tau)-I_{\rm tot}(0)-\sum_{j=A,B}D[\rho_j(0)||\gamma_j^R], & \mathrm{when}~S_C(0)\ge I_{\rm tot}(0),\\
\Delta I_{\rm tot}(\tau)-\sum_{j=A,B}D[\rho_j(0)||\gamma_j^R], &\mathrm{otherwise}.
\end{array}\right.
\end{equation}
\end{widetext}
This inequality, generalizing Eqs. (\ref{eq:catalyst_final}) and (\ref{eq:34}), clearly indicates that an effective ACF can occur in the presence of quantum catalyst as it permits negative lower bounds regardless of the sign of $S_C(0)-I_{\rm tot}(0)$.

\subsubsection{A special case: $\rho_j(0)=\gamma_j=\gamma_j^R$}
We observe that correlated initial states with local marginals given by exact generalized Gibbsian states [cf. Eq. (\ref{eq:ggs})] constitute a special case of our general framework defined by Eqs. (\ref{eq:31})-(\ref{eq:38}). Here, the subsystems' initial states coincide with their reference states $\rho_j(0)=\gamma_j=\gamma_j^R$, yielding well-defined initial thermodynamic gradients with $\tilde{\lambda}_i^j=\lambda_i^j$ ($j=A,B$). This enables rigorous thermodynamic characterization of ACF. In this special scenario, the quantum relative entropy terms $D[\rho_j(0)||\gamma_j^R]$ involved in Eqs. (\ref{eq:31})-(\ref{eq:38}) vanish, leading to simplified inequalities for ACF identification:

\begin{itemize}
        \item [(i)] For quantum catalyst-free systems, we have\\
        \begin{equation}\label{eq:39}
          \sum_{i=0}(\lambda_i^A-\lambda_i^B)\Delta C_i^A(t)~\ge~\Delta I(t).
         \end{equation}
         Together with Eq. (\ref{eq:multiple_combined inequality}), we thus show that the above form holds in quantum catalyst-free systems with multiple charges (including energy-conserving systems as special cases) regardless of initial correlations. Moreover, Eq. (\ref{eq:39}) demonstrates that ACF can indeed occur through the combined action of both the conventional correlation consumption mechanism and the newly identified drag mechanism.
        \item [(ii)] For quantum catalyst-involved systems, we get\\
        \begin{widetext}
        \begin{equation}\label{eq:40}
          \sum_{i=0}(\lambda_i^A-\lambda_i^B)\Delta C_i^A(\tau)~\ge~\left\{\begin{array}{ll}
           2S_C(0)-I_{\rm tot}(\tau)-I_{\rm tot}(0), &\mathrm{when}~S_C(0)\ge I_{\rm tot}(0),\\
           \Delta I_{\rm tot}(\tau), & \mathrm{otherwise}.
         \end{array}\right.
         \end{equation}
         \end{widetext}
         From the above equation, it is interesting to highlight that the choice of inequality for analyzing ACF depends critically on the relative magnitudes of two quantities: (1) the von Neumann entropy of the quantum catalyst $S_C(0)$ and (2) the initial correlations $I_{\rm tot}(0)$ present in the composite system. However, we emphasize that the lower bounds in Eq. (\ref{eq:40}) remain independent of $S_C(0)$, as evident by combining the relation $S_C(\tau)=S_C(0)$ for quantum catalyst and the definition of multipartite quantum mutual information.
\end{itemize}

Allowing for initial correlated states, our results in Eqs. (\ref{eq:39}) and (\ref{eq:40}) provide a straightforward extension of previous studies~\cite{Dijkstra.10.PRL,Semin.12.PRA,Chaudhry.13.PRA,Zhang.15.SR,Jevtic.15.PRE,Ahmadi.21.SR,Colla.22.NJP,Henao.18.PRE,Latune.19.PRR,Medina.20.PRA,Xian.20.PRR,Zicari.20.PRR,Pusuluk.21.PRR,Gnezdilov.23.PRA,Bartosik.24.PRL,Ma.25.PRR,Comar.24.A,Mallik.24.A,Micadei.19.NC} which just considered energy-conserving systems. Importantly, Eq. (\ref{eq:40}), when applying to energy-conserving systems, provides the missing theoretical foundation that was absent in Ref. \cite{Bartosik.24.PRL}, which only preformed numerical simulations of quantum catalyst-involved systems without deriving an associated thermodynamic inequality for the figure of merit $(\beta_A-\beta_B)\Delta E_A(\tau)$. By comparing the lower bounds in Eqs. (\ref{eq:39}) and (\ref{eq:40}), we establish a framework to quantitatively assess whether quantum catalysts can enhance ACF~\cite{Bartosik.24.PRL}.

%==============================================================
\section{Discussion and conclusion}\label{sec:conclusion}
In this work, we developed a feasible global-local thermodynamic description to systematically investigate anomalous transport phenomena of conserved charges in arbitrary correlated quantum systems, aiming to address whether AEF can occur without correlation depletion, or if analogous anomalous transport exists for conserved charges beyond energy. We first rigorously proved a no-go result demonstrating that AEF cannot occur in initially uncorrelated, energy-conserving quantum systems--even when assisted by quantum catalysts. To circumvent this no-go result, we then turned to transport setups with multiple conserved charges and demonstrated how normal flows of non-energy charges can drag AEF without correlation consumption, establishing an alternative paradigm beyond the conventional correlation-involved AEF mechanism. Furthermore, we generalized the notion of AEF to the more general concept of ACF, which provides a universal framework describing counter-gradient transport of arbitrary conserved charges. We confirmed our theoretical expectations with clear numerical evidence. Finally, we extended our framework to incorporate charge non-conservation effects and initially correlated states with nonequilibrium local marginals. These generalizations offer physical insights that further complement existing studies. Our results thus lay down a general theoretical framework for understanding and realizing anomalous flows in correlated quantum systems. 

From an experimental standpoint, the identified drag-induced AEF and its generalized ACF facilitate direct verifications as they eliminate the need for preparing initial correlated states as required by conventional AEF, which often poses significant experimental challenges in quantum systems of a multipartite nature~\cite{Pocklington.25.PRL,Lloyd.25.PRXQ}. Instead,
their implementations require engineering initial thermodynamic gradients through parameter adjustments, thereby enabling the harnessing of normal flows of auxiliary charges to drag the desired anomalous transport. Looking forward to future extensions, we envision applying the current thermodynamic framework and findings to more complex quantum systems, including many-body ones~\cite{Rigol.07.PRL,Langen.15.S} or quantum transport networks~\cite{Maier.19.PRL}, and exploring the anomalous transport behaviors which may lead to design principles of quantum thermal machines. 

As a final remark, we emphasize that although our numerical examples consider minimal models with two commuting charges, the theoretical framework developed in Secs. \ref{sec:multiple} and \ref{sec:extension} B is completely general and applies equally to noncommuting charges. This universality arises because the initial generalized Gibbs states--which underlie our key expressions for multiple charges--remain rigorously defined even for noncommuting cases~\cite{Halpern.16.NC,Guryanova.16.NC}. Nevertheless, experimental implementation presents significant challenges in such scenarios, as noncommutativity fundamentally restricts simultaneous measurement of all charge variations. These intriguing possibilities remain open for future investigation.

\section*{Acknowledgments}
J.L. acknowledges support from the National Natural Science Foundation of China (Grant No. 12205179), the Shanghai Pujiang Program (Grant No. 22PJ1403900) and start-up funding of Shanghai University.

\appendix
\renewcommand{\theequation}{A\arabic{equation}}
\renewcommand{\thefigure}{A\arabic{figure}}
\setcounter{equation}{0}  % reset counter
\setcounter{figure}{0}  % reset counter
\section{Derivation details}
\label{a:1}
In this appendix, we present the derivation details of several key expressions shown in the main text. Firstly, we briefly recall some definitions that facilitate the subsequent derivations. The von Neumann entropy of a system with density matrix $\rho$ is defined as 
\bea
S(\rho)=-\text{Tr}[\rho \ln \rho].
\eea
Moreover, the quantum relative entropy is given by 
\bea
D[\rho_1||\rho_2] = \text{Tr}[\rho_1(\ln \rho_1 -\ln \rho_2)].
\eea
Finally, the quantum mutual information between two subsystems is defined as
\bea
I(t) = S_A(t)+S_B(t)-S(t)=D[\rho_{AB}||\rho_A \otimes \rho_B].
\eea
where $S_i(t)$ ($i=A,B$) denote the von Neumann entropy with respect to the reduced density matrix of subsystem $i$ $\rho_i=\text{Tr}_{\tilde{i}}[\rho]$ by tracing out degrees of freedom $\tilde{i}$ that complements $i$, and $S(t)$ is the von Neumann entropy of the total system. 

\subsection{Entropy change under initial thermal states}
We begin by detailing the procedures used to evaluate the change in the von Neumann entropy of a subsystem shown in Eq. (\ref{eq:dsa}) of the main text, following closely the information-theoretic approach introduced in Ref. \cite{Esposito.10.NJP} for deriving the second law of thermodynamics in system-bath configurations. Starting from an initial product state where each subsystem is prepared in its thermal state, the bipartite system evolves under an energy-conserving unitary transformation as elaborated in the main text. In general, the change in the von Neumann entropy of subsystem $A$ during a time interval $[0,t]$ reads
\bea \label{eq:decomposition}
\Delta S_A(t) &\equiv& S_A(t) - S_A(0)\nonumber\\
&=& -\text{Tr}[\rho_A(t)\ln \rho_A(t)] +  \text{Tr}[\rho_A(0)\ln \rho_A(0)]\nonumber\\
&=& -\text{Tr}[\rho(t)\ln \rho_A(t)] +  \text{Tr}[\rho_A(0)\ln \rho_A(0)].
\eea
To get the third line, we have utilized the reduced state definition $\rho_A(t)=\mathrm{Tr}_B[\rho(t)]$ by tracing out degrees of freedom of subsystem $B$. Hence, the trace operation in the first term of the third line is performed with respect to the total system, whereas trace operations in other terms is taken over degrees of freedom of solely subsystem $A$. As the subsystems are uncorrelated at the initial moment with $\rho(0)=\rho_A(0)\otimes\rho_B(0)$, the von Neumann entropy of the total system $S(0)=-\text{Tr}[\rho(0)\ln \rho(0)]$ can be decomposed as $S(0) = S_{A}(0) + S_{B}(0)$. We then invoke this initial entropy decomposition to replace the term $\text{Tr}[\rho_A(0)\ln \rho_A(0)]=-S_A(0)$ in Eq. (\ref{eq:decomposition}), yielding
\bea\label{eq:a5}
\Delta S_A(t) &=& -\text{Tr}[\rho(t)\ln \rho_A(t)] + \text{Tr}[\rho(0)\ln \rho(0)]\nonumber \\
   &&- \text{Tr}[\gamma_B\ln \gamma_B]+ \text{Tr}[\rho_B(t)\ln \gamma_B]\nonumber \\
   &&-\text{Tr}[\rho_B(t)\ln \gamma_B].
\eea
Here, we have utilized the fact that $\rho_B(0)=\gamma_B$ such that $S_B(0)=-\text{Tr}[\gamma_B\ln \gamma_B]$. We note the last two terms on the right-hand side of Eq. (\ref{eq:a5}) cancel out, making the equality unchanged.

As the composite system undergoes a unitary evolution, the total von Neumann entropy remains invariant, namely $S(0)=S(t)$. We can then rearrange Eq. (\ref{eq:a5}) into 
\bea\label{eq:a6}
\Delta S_A(t) &=& -\text{Tr}[\rho(t)\ln (\rho_A(t) \otimes \gamma_B)] + \text{Tr}[\rho(t)\ln \rho(t)]\nonumber\\
&& + \text{Tr}[(\rho_B(t) - \gamma_B)\ln \gamma_B]\nonumber\\
   &=& D[\rho(t)||\rho_A(t) \otimes \gamma_B] - \beta_B\Delta E_B(t).
\eea
To get the first term on the right-hand side of Eq. (\ref{eq:a6}), we have noted that $\text{Tr}[\rho_B(t)\ln \gamma_B]=\text{Tr}[\rho(t)\ln \gamma_B]$. It is evident that Eq. (\ref{eq:a6}) is just Eq. (\ref{eq:dsa}) of the main text. From the above derivation, we remark that Eq. (\ref{eq:a6}) involves no approximations besides an initial product thermal state requirement, thus representing a quite general expression for subsystems in any bipartite correlated quantum systems.

We then turn to setups with quantum catalyst. The generalization of Eq. (\ref{eq:a6}) is rather straightforward. One just needs to consider an enlarged composite system with an initial product state
$\rho_{\rm tot}(0) = \rho(0) \otimes \omega_C$, where $\rho(0) = \gamma_A \otimes \gamma_B$ and $\omega_C$ denotes the initial state of catalyst that is determined self-consistently according to Eq. (\ref{eq:cata_state}) of the main text. Starting from Eq. (\ref{eq:decomposition}), we find
\bea\label{eq:a7}
   \Delta S_A(t) &=& -\text{Tr}[\rho_A(t)\ln \rho_A(t)] +  \text{Tr}[\rho_A(0)\ln \rho_A(0)]\nonumber\\
   &=& -\text{Tr}[\rho_{\rm tot}(t)\ln \rho_A(t)] + \text{Tr}[\rho_{\rm tot}(0)\ln \rho_{\rm tot}(0)]\nonumber \\
   &&- \text{Tr}[\gamma_B\ln \gamma_B] - \text{Tr}[\omega_C\ln \omega_C]\nonumber \\
   &&+ \text{Tr}[\rho_B(t)\ln \gamma_B] -\text{Tr}[\rho_B(t)\ln \gamma_B]\nonumber\\
   &=& -\text{Tr}[\rho_{\rm tot}(t)\ln (\rho_A(t) \otimes \gamma_B\otimes \mathrm{I}_C))]\nonumber\\
   &&+ \text{Tr}[\rho_{\rm tot}(t)\ln \rho_{\rm tot}(t)] + \text{Tr}[(\rho_B(t) - \gamma_B)\ln \gamma_B]\nonumber\\
   &&-\text{Tr}[\omega_C\ln \omega_C]\nonumber\\ 
   &=& D[\rho_{\rm tot}(t)||\rho_A(t) \otimes \gamma_B\otimes \mathrm{I}_C] - \beta_B\Delta E_B(t)\nonumber \\
   &&+S_C(0).
\eea
To arrive at the second line, we have utilized the decomposition of initial total von Neumann entropy $S_{\rm tot}(0) = S_{A}(0)+S_{B}(0)+S_C(0)$ to replace the term $\text{Tr}[\rho_A(0)\ln \rho_A(0)]$ in the first line. To get the third equality, we have invoked the relations $\text{Tr}[\rho_{\rm tot}(0)\ln \rho_{\rm tot}(0)]=\text{Tr}[\rho_{\rm tot}(t)\ln \rho_{SC}(t)]$ due to the invariance of total von Neumann entropy under unitrary transformation, and $\text{Tr}[\rho_B(t)\ln \gamma_B]=\text{Tr}[\rho_{\rm tot}(t)\ln \gamma_B]$. We have further introduced $\mathrm{I}_C$ to denote the identity matrix of the dimension of the catalyst. Eq. (\ref{eq:a7}) is just Eq. (\ref{eq:entropy_catalyst}) of the main text.

Extending Eq. (\ref{eq:a6}) to multiple-charge scenarios is also straightforward: Noting that one considers a generalized Gibbsian state as the initial state, one just needs to replace the contribution $\beta_B\Delta E_B(t)$ with $\sum_{i=0}\lambda_i^{B}\Delta C_i^{B}(t)$ in Eq. (\ref{eq:a6}) such that one gets Eq. (\ref{eq:entropy_multiple}) shown in the main text. 

\subsection{Global-local transformation of quantum relative entropy}
Here we focus on deriving the global-local transformation formula utilized in the main text to get a local description from a global one. Starting from quantum relative entropy $D[\rho(t)||\rho_A(t) \otimes \gamma_B]$ involving the global state $\rho(t)$, we can split it as 
\bea\label{eq:a8}
D[\rho(t)||\rho_A(t) \otimes \gamma_B] &=& \text{Tr}[\rho(t) \ln \rho(t)] \nonumber \\
&&- \text{Tr}[\rho(t) \ln( \rho_A(t) \otimes \gamma_B)]\nonumber\\
   &=& \text{Tr}[\rho(t) \ln \rho(t)]-\text{Tr}[\rho_A(t)\ln \rho_A(t)]\nonumber\\
   &&-\text{Tr}[\rho_B(t)\ln \gamma_B].
\eea
Utilizing the definition of quantum mutual information, we can further combine the first and second terms on the right-hand side of the last equality in Eq. (\ref{eq:a8}), yielding 
\bea
D[\rho(t)||\rho_A(t) \otimes \gamma_B] &=& I(t) + \text{Tr}[\rho_B(t)\ln \rho_B(t)]\nonumber\\
&&-\text{Tr}[\rho_B(t)\ln \gamma_B].
\eea
Notably, the last two terms together just correspond to a quantum relative entropy between states $\rho_B(t)$ and $\gamma_B$
\bea\label{eq:a10}
 D[\rho(t)||\rho_A(t) \otimes \gamma_B] = I(t) + D[\rho_B(t)||\gamma_B].
\eea
Here, we regard $D[\rho_B(t)||\gamma_B]$ as local quantum relative entropy because it only involves the local density matrix of the subsystem. Thus, we obtain the global-local transformation expression used in the main text. We point out that Eq. (\ref{eq:a10}) can be directly applied to setups with multiple charges.

As for the scenario with catalyst, we focus on $D_{\rm tot}\equiv D[\rho_{\rm tot}(t)||\rho_A(t) \otimes \gamma_B\otimes \mathrm{I}_C]$ which can be proceeded as follows
\bea\label{eq:a11}
   D_{\rm tot} &=& \text{Tr}[\rho_{\rm tot}(t) \ln \rho_{\rm tot}(t)]- \text{Tr}[\rho_{\rm tot}(t) \ln( \rho_A(t) \otimes \gamma_B)]\nonumber\\
   &=& \text{Tr}[\rho_{\rm tot}(t) \ln \rho_{\rm tot}(t)]-\text{Tr}[\rho_A(t)\ln \rho_A(t)]\nonumber\\
   &&-\text{Tr}[\rho_B(t)\ln \gamma_B]\nonumber\\
   &=& I_{\rm tot}(t) -S_C(t) + \text{Tr}[\rho_B(t)\ln \rho_B(t)]\nonumber\\
   &&-\text{Tr}[\rho_B(t)\ln \gamma_B]\nonumber\\
   &=& I_{\rm tot}(t) -S_C(t)+ D[\rho_B(t)||\gamma_B].
\eea
Here, $I_{\rm tot}(t)=S_A(t)+ S_B(t)+S_C(t)-S_{\rm tot}(t)$ represents a multiple generalization of the quantum mutual information~\cite{Groisman.05.PRA,Modi.12.RMP}. This equality is the global-local transformation utilized in the main text for analyzing setups with catalyst. 

\subsection{Generalization to nonequilibrium initial states}
Lastly, we generalize Eqs. (\ref{eq:a5})-(\ref{eq:a6}) to account for non-Gibbsian initial states $\rho_{A,B}(0)$ by introducing initial reference states defined in Eq. (\ref{eq:reference_state}). Starting from Eq. (\ref{eq:decomposition}) which remains applicable, we generalize Eq. (\ref{eq:a5}) to obtain
\bea\label{eq:a12}
\Delta S_A(t) &=& -\text{Tr}[\rho(t)\ln \rho_A(t)] + \text{Tr}[\rho(0)\ln \rho(0)]\nonumber \\
&&-\text{Tr}[\rho_B(0)\ln \rho_B(0)]-I(0)\nonumber\\
&=& -\text{Tr}[\rho(t)\ln \rho_A(t)] + \text{Tr}[\rho(0)\ln \rho(0)]\nonumber \nonumber\\
&&-\text{Tr}[\rho_B(0)\ln \rho_B(0)]-I(0)\nonumber\\
   &&-\text{Tr}[\rho_B(0)\ln \gamma_B^R]+ \text{Tr}[\rho_B(0)\ln \gamma_B^R]\nonumber \nonumber\\
   &&-\text{Tr}[\rho_B(t)\ln \gamma_B^R]+ \text{Tr}[\rho_B(t)\ln \gamma_B^R].
\eea
Here, in arriving at the first equality, we have utilized the definition of initial quantum mutual information $I(0)=S_A(0)+S_B(0)-S(0)$ to rewrite the initial entropy of subsystem $A$ in Eq. (\ref{eq:decomposition}) and noted that $\rho_B(0)$ is no longer thermal. Noted that we do not necessarily require $I(0)=0$ here. By combining terms in the last equality, Eq. (\ref{eq:a12}) becomes
\bea\label{eq:a13}
\Delta S_A(t) &=& -\text{Tr}[\rho(t)\ln (\rho_A(t) \otimes \gamma_B^R)] + \text{Tr}[\rho(t)\ln \rho(t)]\nonumber\\
&& -D[\rho_B(0)||\gamma_B^R]-I(0)\nonumber\\
&& + \text{Tr}[(\rho_B(t) - \rho_B(0))\ln \gamma_B^R]\nonumber\\
   &=& D[\rho(t)||\rho_A(t) \otimes \gamma_B^{R}] - \sum_{i=0}\tilde{\lambda}_i^B\Delta C_i^B(t)\nonumber\\
   &&-D[\rho_B(0)||\gamma_B^R]-I(0).
\eea
This is just Eq. (\ref{eq:31}) in the main text. In arriving at the first line of the above equation, we have utilized the fact that the total von Neumann entropy remains invariant under unitary evolution, namely $S(0)=S(t)$. In getting the second equality of the above equation, we have utilized the form of generalized Gibbsian reference state in Eq. (\ref{eq:reference_state}). We also note that Eq. (\ref{eq:a10}) can be generalized to 
\begin{equation}
 D[\rho(t)||\rho_A(t) \otimes \gamma_B^R] = I(t) + D[\rho_B(t)||\gamma_B^R].   
\end{equation}

%--------------------------------------------------------
\begin{table}[t!]
\centering
\caption{Summary of main notations}
\begin{tabularx}{\columnwidth}{p  {1.2 cm}   l }
\hline
\hline
$\rho_j$  & Reduced state of subsystem $j$ \\
$\gamma_j$ & Gibbisian/Generalized Gibbsian states of subsystem $j$ \\
$\rho$ & State of bipartite systems \\
$\rho_{\rm tot}$ & State of composite systems with catalyst \\
$S_j$  & von Neumann entropy of subsystem $j$ \\ 
$S$ & von Neumann entropy of bipartite systems \\
$S_{\rm tot}$ & von Neumann entropy of composite systems with catalyst \\
$\mathcal{C}_i^j$  & $i$th charge of subsystem $j$\\
$\lambda_i^j$  & Thermodynamic affinity for $i$th charge of subsystem $j$ \\ 
$\mathcal{L}_{g,l}$ & \makecell[tl]{Lower bounds on $(\beta_A-\beta_B)\Delta E_A$ in energy-conserving \\ system with catalyst}\\
$\mathcal{L}_{g,l}^E$ & \makecell[tl]{Lower bounds on $(\beta_A-\beta_B)\Delta E_A$ in systems with multiple \\ charges} \\
$\mathcal{L}_{g,l}^C$ & \makecell[tl]{lower bounds on $(\lambda_k^A-\lambda_k^B)\Delta C_k^A$ in systems with multiple \\ charges} \\
$\omega_C$ & Quantum catalyst's state\\
$D[\rho_1||\rho_2]$ & Quantum relative entropy between states $\rho_{1,2}$\\
$I$ & Quantum mutual information of bipartite systems\\
$I_{\rm tot}$ & \makecell[tl]{Quantum mutual information of composite systems with \\ catalyst}\\
\hline
\hline
\end{tabularx}
\label{table1}
\end{table}
%--------------------------------------------------------
In the presence of a quantum catalyst, we should generalize Eq. (\ref{eq:a7}) to account for nonequilibrium initial states,
\bea\label{eq:a7}
   \Delta S_A(t) &=& -\text{Tr}[\rho_A(t)\ln \rho_A(t)] +  \text{Tr}[\rho_A(0)\ln \rho_A(0)]\nonumber\\
   &=& -\text{Tr}[\rho_{\rm tot}(t)\ln \rho_A(t)] + \text{Tr}[\rho_{\rm tot}(0)\ln \rho_{\rm tot}(0)]\nonumber \\
   &&- \text{Tr}[\rho_B(0)\ln \rho_B(0)] - \text{Tr}[\omega_C\ln \omega_C]-I_{\rm tot}(0)\nonumber \\
   &&+\text{Tr}[\rho_B(0)\ln \gamma_B^R]-\text{Tr}[\rho_B(0)\ln \gamma_B^R]\nonumber\\
   &&+\text{Tr}[\rho_B(t)\ln \gamma_B^R] -\text{Tr}[\rho_B(t)\ln \gamma_B^R]\nonumber\\
   &=& -\text{Tr}[\rho_{\rm tot}(t)\ln (\rho_A(t) \otimes \gamma_B^R\otimes \mathrm{I}_C))]\nonumber\\
   &&+ \text{Tr}[\rho_{\rm tot}(t)\ln \rho_{\rm tot}(t)] + \text{Tr}[(\rho_B(t) - \rho_B(0))\ln \gamma_B^R]\nonumber\\
   &&-\text{Tr}[\omega_C\ln \omega_C]-I_{\rm tot}(0)-D[\rho_B(0)||\gamma_B^R]\nonumber\\ 
   &=& D[\rho_{\rm tot}(t)||\rho_A(t) \otimes \gamma_B^R\otimes \mathrm{I}_C]+S_C(0)-I_{\rm tot}(0) \nonumber \\
   &&-D[\rho_B(0)||\gamma_B^R]-\sum_{i=0}\tilde{\lambda}_i^B\Delta C_i^B(t).
\eea
To get the second equality, we have utilized the definition of $I_{\rm tot}(0)$ to replace the initial von Neumann entropy of subsystem $A$. In arriving at the last equality which is just Eq. (\ref{eq:35}) in the main text, we have used the definition $S_C(0)=-\text{Tr}[\omega_C\ln \omega_C]$ and the form of generalized Gibbsian reference state in Eq. (\ref{eq:reference_state}). Similar to Eq. (\ref{eq:a11}), we have $D[\rho_{\rm tot}(t)||\rho_A(t) \otimes \gamma_B^R\otimes \mathrm{I}_C]=I_{\rm tot}(t) -S_C(t)+ D[\rho_B(t)||\gamma_B^R]$. By swapping the indices $A\leftrightarrow B$, we can obtain an expression for $D[\rho_{\rm tot}(t)||\rho_B(t) \otimes \gamma_A^R\otimes \mathrm{I}_C]$.

For reader convenience, we provide a summary of main notations used in this study in Table \ref{table1}.

%\bibliography{AEF}


\end{document}